\documentclass[fleqn,10pt]{wlscirep}
\title{Symmetry breaking,  Josephson oscillation and self-trapping in a self-bound three-dimensional quantum ball}

\author[1,*]{S. K. Adhikari}
\affil[1]{Instituto de F\'{\i}sica Te\'orica, UNESP - Universidade Estadual Paulista, 01.140-070 S\~ao Paulo, S\~ao Paulo, Brazil}
\affil[*]{adhikari44@yahoo.com}


\begin{abstract}We study spontaneous symmetry breaking (SSB), Josephson oscillation, and self-trapping   in a    stable, mobile,   three-dimensional 
  matter-wave spherical 
quantum ball self-bound by attractive two-body and repulsive three-body interactions. The SSB is realized 
by   a parity-symmetric (a)    one-dimensional (1D) double-well  potential or  (b) a 1D  
Gaussian  potential, both along the $z$ axis  and no potential along the $x$ and $y$  axes. In the presence of each of these 
 potentials, the symmetric ground state dynamically evolves into  a doubly-degenerate  SSB ground 
state. 
If the SSB  ground state in the double well, predominantly located in the first well ($z>0$), is given a small  displacement, the quantum ball oscillates with
  a self-trapping in the first well. For a medium displacement    one encounters  an asymmetric Josephson oscillation. The   asymmetric oscillation is a consequence of SSB.
The study is performed by a variational and numerical solution of a non-linear mean-field model with 
    1D parity-symmetric 
perturbations.  
\end{abstract}
\begin{document}

\flushbottom
\maketitle
\thispagestyle{empty}

%\noindent Please note: Abbreviations should be introduced at the first mention in the main text – no abbreviations lists. Suggested structure of main text (not enforced) is provided below.

\section{Introduction}
The topic of spontaneous symmetry breaking (SSB) in localized quantum states obeying Schr\"odinger dynamics 
has drawn much attention lately both in experimental \cite{symb} and theoretical \cite{m1,m2,m2a,s1a,s1b,s1c} fronts.  
In simple two-boson non-relativistic quantum mechanics,  parity  is an exact symmetry
and the ground state is non-degenerate. 
However, this symmetry can be broken even in non-relativistic quantum mechanics, involving many particles,
bosons \cite{m1,m2,m2a} and  fermions \cite{fermion}. There have been numerous studies of SSB in non-linear optics 
using the non-linear Schr\"odinger (NLS)  equation \cite{m3a,m3b,m3e}.
  There have also been studies of SSB in a localized   Bose-Einstein condensate (BEC)
 \cite{n1a,n1b,n1c}.  The simplest and commonly studied case  of SSB in non-linear systems is an 
attractive one-dimensional (1D) BEC in a confining double-well potential with a doubly
degenerate SSB ground state \cite{m1,m2,s1a,s1b,s1c}. The SSB is found to appear in this system below a threshold of attractive non-linearity or above a 
threshold of the double-well barrier height \cite{m1}. Apart from the studies of SSB in a single-component BEC \cite{m2,m2a}, the other studies on this
topic deal with trapped multi-component BEC \cite{s1a,s1b,s1c}. 
The collapse instability of a two- (2D) or three-dimensional (3D) attractive BEC disappears 
in a 1D BEC \cite{sol}, hence the study of SSB in an attractive 1D BEC should be complimented by a similar study in a 2D and 3D BEC
to verify if the SSB in an attractive 1D BEC does not take place exclusively in the domain of collapse in the      3D configuration.   There has also been study of SSB in two-component trapped  two-dimensional BEC
where the symmetry breaking originating from a phase separation is achieved due to an interplay between   inter-species  
interaction and trapping potential \cite{n2}.   

 The ground state of a {\it repulsive} 1D BEC in a double well is symmetric due to atomic repulsion with an equal number of atoms in both wells. If a small initial population imbalance between the two wells in created, repulsive atoms naturally move back and forth from one well to the another thus initiating a Josephson oscillation  \cite{josep1,josep1a,josep1b,josep1d}.  However, if the initial    population imbalance   is increased in a sufficiently repulsive BEC \cite{st1,st1a,st1b,st1c} or a Fermi super-fluid \cite{stf}, a dynamical self-trapping of the atoms in one well takes place resulting in a net time-averaged population imbalance. The self-trapping of a larger number of atoms in one well compared to the other is counter-intuitive in a repulsive BEC.  
  
Motivated by the above consideration, in this paper  we study  SSB, Josephson oscillation \cite{josep1,josep1a,josep1b,josep1d}  and self trapping  
in a  3D self-bound {\it attractive} matter-wave  quantum ball \cite{qb1,qb2,qb3} placed in  a parity-symmetric (a)    1D double-well  potential or  (b) a 1D  
Gaussian   potential along the $z$ axis. These 1D potentials are necessary for SSB; however, these  potentials act in 1D and thus 
have no effect on the localization of the 3D quantum ball. 
 The 3D quantum ball is stabilized in the presence of   a repulsive three-body interaction  and an adequate two-body attraction above a critical value  \cite{qb1,qb2,qb3}.
It has been demonstrated that a tiny three-body repulsion stops the collapse and  can stabilize a 3D quantum ball for zero \cite{qb1} and non-zero angular momenta \cite{qb4}.  This gives the unique opportunity to study SSB 
in a 3D quantum ball bound solely by non-linear interactions placed in these parity-symmetric 1D potentials. 
For the first time we detect an asymmetric Josephson oscillation because of the SSB in the quantum ball in the presence of these two 1D potentials.    The quantum ball exhibits SSB in these potentials $-$ the symmetric ground state   is displaced from the origin along
 the $z$ axis as a consequence of SSB, thus violating the parity symmetry of the   Hamiltonian  and leading  to a doubly-degenerate ground state. The quantum ball can be displaced along  either the positive or negative $z$ axis leading to two degenerate states.   
We find that  the SSB 
occurs  for the weakest possible   double-well  or Gaussian potential.

The SSB takes place in an attractive BEC due to a sizable nonlinear interaction. Previous studies of SSB in trapped attractive BEC \cite{m1,m2,m2a} have employed a moderately large nonlinearity in a quasi-1D configuration.  For such values of attractive nonlinearity, the corresponding trapped 3D BEC collapses, whereas the reduced 1D model does not collapse. In the present study, collapse has been stopped by a three-body repulsion and an adequate two-body attraction.

% The predominantly attractive 3D matter-wave  quantum ball in a 1D double-well potential provides a new scenario for the study of  Josephson oscillation and self-trapping.

 Unlike in a trapped repulsive BEC in a double-well potential with zero steady-state population imbalance, the population imbalance in a SSB ground state of a quantum ball in a double well is usually large. If the population imbalance  in a SSB ground state, predominantly located in the first well ($z>0$),   is changed   by displacing the ground state towards or away from the center of the double well ($z=0$) by a small  distance $\delta$,  a small oscillation of the quantum ball indicating self-trapping in the first well results.   If the initial displacement is large and away from the trap center, one has a symmetric Josephson oscillation between the two wells.  For a medium  initial displacement towards or away from the trap center, an asymmetric Josephson oscillation between the two wells take place: two extreme states 
of the    oscillating  SSB BEC are not parity image of each other. This manifestly asymmetric Josephson oscillation is a consequence of SSB in the quantum ball in the presence of 1D double-well potential. 
For larger initial displacement towards the center of the trap, a self-trapping in the second well results.  
 In the present case of the attractive quantum ball, the self trapping in the first and the second well for small and large initial displacements is not quite surprising, but the continued asymmetric  Josephson oscillation of most atoms for a medium displacement is counter-intuitive as atomic attraction should permanently take all the atoms to one of the wells.      
The SSB also manifests when the self-bound quantum ball is placed 
on the top of a  parity-symmetric 1D Gaussian potential hill. Like a classical ball the   quantum ball is found to slide down the potential hill, thus spontaneously breaking the symmetry.

We base the present study on a variational approximation and a numerical solution of the 3D  mean-field
Gross-Pitaevskii (GP) equation in the presence of an  attractive  two-body and repulsive three-body interactions.   The two-body contact attraction leads to a cubic non-linear term in the GP equation and an attractive  cubic divergence near the origin in the effective Lagrangian, viz. Eq. (\ref{eq4}), responsible for collapse instability. 
The three-body contact repulsion, on the other hand, leads to a quintic non-linearity and a repulsive  sextic divergence near the origin suppressing the attractive  cubic divergence, thus stopping the collapse.

The mathematical structure of the non-linear mean-field  GP equation is the same as that of the NLS equation used to study pulse propagation in non-linear optics, although the physical meaning of the different terms is distinct in two cases.   Hence,   in a cubic-quintic non-linear medium
\cite{cq,cqb,cqc} one can have a stable mobile 3D spatiotemporal light bullet \cite{qb3,qb4} and a SSB can occur in that context also.

We present the 3D GP  equation used in this study  in Sec. 
\ref{II} and an analytic variational approximation to it. 
In Sec. \ref{III} we present the numerical and variational results for stationary profiles of  SSB quantum ball under  perturbations in the form of a 1D double-well  or a 1D  Gaussian  potential. 
We study how a parity-symmetric state dynamically evolves into a SSB state under the action of the perturbation.     We study Josephson oscillation and self-trapping  of the SSB quantum ball in the 1D double-well potential. A description of the numerical methods for the solution of the GP equation is given in Sec. \ref{IV}.
We end with a summary and discussion in Sec. \ref{V}.

\section{Result}

\subsection{The GP equation and   Variational approximation}
 
\label{II}
  
The mean-field GP equation describing the BEC  quantum ball 
in the presence of an attractive two-body and a repulsive three-body interaction  subject to an external perturbation $V(z)$
 is given by \cite{qb1,qb2}
\begin{align}\label{eq31}
i\hbar  \frac{\partial \phi({\bf r},t)}{\partial t}  &=\biggr[-\frac{\hbar^2}{2m} \nabla^2_{\bf r} +V(z)-\frac{4\pi  |a| \hbar^2N}{m}|  \phi({\bf r},t) |^2 
 +\frac{\hbar N^2 K_3}{2} |  \phi({\bf r},t) |^4  \biggr]  \phi({\bf r},t), \\
V(z) &= \frac{c_1}{2}m\omega^2 z^2 
+{c_2}\hbar \omega e^{-\gamma m\omega z^2/\hbar},\label{dw}
 \end{align}
where  
$m$ is the mass of each atom,  $\phi({\bf r},t)$ is the condensate wave function at a space point ${\bf r}=\{x,y,z\}$ and time $t$,   $a$ is the $s$-wave scattering length of atoms taken here to be negative (attractive), $K_3$ is the three-body interaction term, and $N$ is the number of atoms.
The external double-well potential  $V(z)$ consists of a harmonic potential of strength $c_1$ and angular frequency $\omega$ and a Gaussian potential of strength $c_2$ and width parameter $\gamma$. If $c_1$ is set zero, this potential becomes a Gaussian potential. 
 Equation (\ref{eq31}) can be written in the following dimensionless form after a redefinition of the variables
\begin{align}\label{eq2}
i\frac{\partial \phi({\bf r},t)}{\partial t}   =\biggr[-\frac { \nabla^2_{\bf r}}{2} +\frac{c_1}{2}z^2+{c_2}e^{-\gamma z^2}  - p |\phi({\bf r},t) |^2  
   &+q|  \phi({\bf r},t) |^4  \biggr]  \phi({\bf r},t),
 \end{align}
where 
$p=4\pi Na/l$, $q= m N^2 K_3/(2\hbar l^4)$, length is scaled in units of $l\equiv \sqrt{\hbar/{m\omega}}$, time in $ml^2/\hbar$, $|\phi|^2$ in units of $l^{-3}$, and energy in units of $\hbar \omega$. 
For a stationary state with property $\phi({\bf r},t)\sim \phi({\bf r})\exp(-i\mu t)$, with $\mu$ the chemical potential,
   one has the following time-independent GP equation:
  \begin{align}\label{eq4}
\mu    \phi({\bf r})=\biggr[-\frac { \nabla^2_{\bf r}}{2} +\frac{c_1}{2}z^2+{c_2}e^{-\gamma z^2}  - p |\phi({\bf r}) |^2  
   &+q|  \phi({\bf r}) |^4  \biggr]  \phi({\bf r}),
 \end{align}

 For an analytic understanding of SSB of a quantum ball,
we consider the   Lagrange  variational formulation  \cite{var} to Eq. (\ref{eq4}). 
 In this axially symmetric problem, convenient analytic Gaussian 
 variational approximation of the  quantum ball wave function  is   \cite{var}
\begin{align}\label{eq3}
 \phi({\bf r})&=\frac{ \pi^{-3/4}}{\sigma_1{\sigma_2^{1/2}}}\exp\biggr[-\frac{\rho^2}{2\sigma_1^2}- \frac{(z-z_0)^2}{2\sigma_2^2}\biggr], %\nonumber \\
% &+ i \alpha(t) \rho^2+i \beta(t)   z^2\biggr],
\end{align}
where   $\rho=\sqrt{x^2+y^2},  \sigma_1$ and $\sigma_2$ are radial and axial  widths, respectively. In this approximation, the Gaussian in the $z$ direction is displaced by a distance $z_0$ from the origin due to  the SSB under perturbation $V(z)$. The $z$-profile of the displaced SSB quantum ball is not strictly a Gaussian, but is close to it, as we will see, and here for simplicity we take it as a Gaussian.
 The (generalized) Lagrangian  density corresponding to Eq. (\ref{eq4}) is  
\begin{align}\label{eq41}
{\cal L}({\bf r}) = 
\frac{|\nabla \phi({\bf r}) |^2}{2}
+\frac{c_1}{2}z^2  | \phi({\bf r})|^2 &+{c_2}e^{-\gamma z^2}  | \phi({\bf r})|^2 -\frac{p}{2}| \phi({\bf r})|^4
+\frac{q}{3}| \phi({\bf r})|^6.
\end{align} 
Equation (\ref{eq4}) can be obtained by extremizing the functional (\ref{eq41}) \cite{var}. 
Consequently, the effective Lagrangian functional $\overline L(\sigma_1,\sigma_2,z_0) \equiv 2\pi \int {\cal L}({\bf r}) dz \rho d\rho$ becomes
\begin{align}\label{eq6}
\overline L =  
\frac{1}{2\sigma_1^2}+\frac{1}{4\sigma_2^2}+
 \frac{c_1}{4}   \left( \sigma_2 ^2+ 2{z_0^2} \right)
+\frac{c_2 e^{-\frac{ \gamma z_0^2}{1+\gamma \sigma_2^2}}}{\sqrt{1+\gamma \sigma_ 2^2}}
-\frac{p\pi^{-3/2}}{4\sqrt 2  \sigma_1^2\sigma_2}
+\frac{q\pi^{-3}}{9\sqrt 3  \sigma_1^4 \sigma_2^2}. 
\end{align}   
Lagrangian (\ref{eq6}) is the energy per atom in the SSB quantum ball. 
The variational parameters $\nu\equiv \sigma_1, \sigma_2, z_0$ are obtained from a minimization 
of the effective Lagrangian functional $\overline L: $
\begin{equation}\label{eq7}
   \frac{\partial \overline L}{\partial  \nu}=0.
\end{equation}

\subsection{Numerical Results}

\label{III}
For   illustration in this paper, we consider attractive $^7$Li atoms with scattering length $a=-27.4a_0$ \cite{abr} and a  variable three-body interaction term $K_3$, where $a_0$ is the Bohr radius. The three-body atom loss rate due to  the formation of molecules   is not accurately known \cite{loss}   for relatively low-density quantum balls  in the trap-less domain employed in this study. The effect of this loss rate is expected to be considerable in a large high-density trapped 
attractive BEC and  will   be negligible   for the small-time dynamics (of about 5 ms) of untrapped quantum balls presented in this paper, as was demonstrated in Ref. \cite{qb1}, and hence is not considered here.   
We take the harmonic oscillator length $l=1$ $\mu$m, which corresponds to a trap of angular frequency $\omega = 2\pi \times 1444$ s$^{-1}$,   unit of time $t_0\equiv ml^2/\hbar = 0.11$ ms, and of energy $\hbar \omega = 9.57\times 10^{-31}$ J. The parameters of the double-well potential (\ref{dw}) are taken as $c_1=c_2=1,$ and $\gamma=20$ and of the Gaussian potential as 
$c_1=0,c_2=1,$ and $\gamma=20$.

A stable quantum ball corresponds to a global minimum of  the conserved effective Lagrangian 
$\overline L(\sigma_1,\sigma_2,z_0)$ (\ref{eq6}).   At the center of the $\sigma_1$-$\sigma_2$ plane, $\sigma_1, \sigma_2 \to 0$, and the Lagrangian $\overline L (\sigma_1,\sigma_2,z_0) \to + \infty$, which guarantees the absence of a collapsed state at the origin. The statics and the dynamics of a self-bound 3D quantum ball in the absence of the 1D double well ($c_1=c_2=0$) have been studied in details \cite{qb1,qb2}. The quantum ball is bound for any value of the quintic non-linearity $q$ and for the cubic non-linearity $p$ above a critical
 value  $p>p_{\mathrm{crit}}$ \cite{qb1,qb2}.

\begin{figure}[!t]

\begin{center}
\includegraphics[width=.45\linewidth,clip]{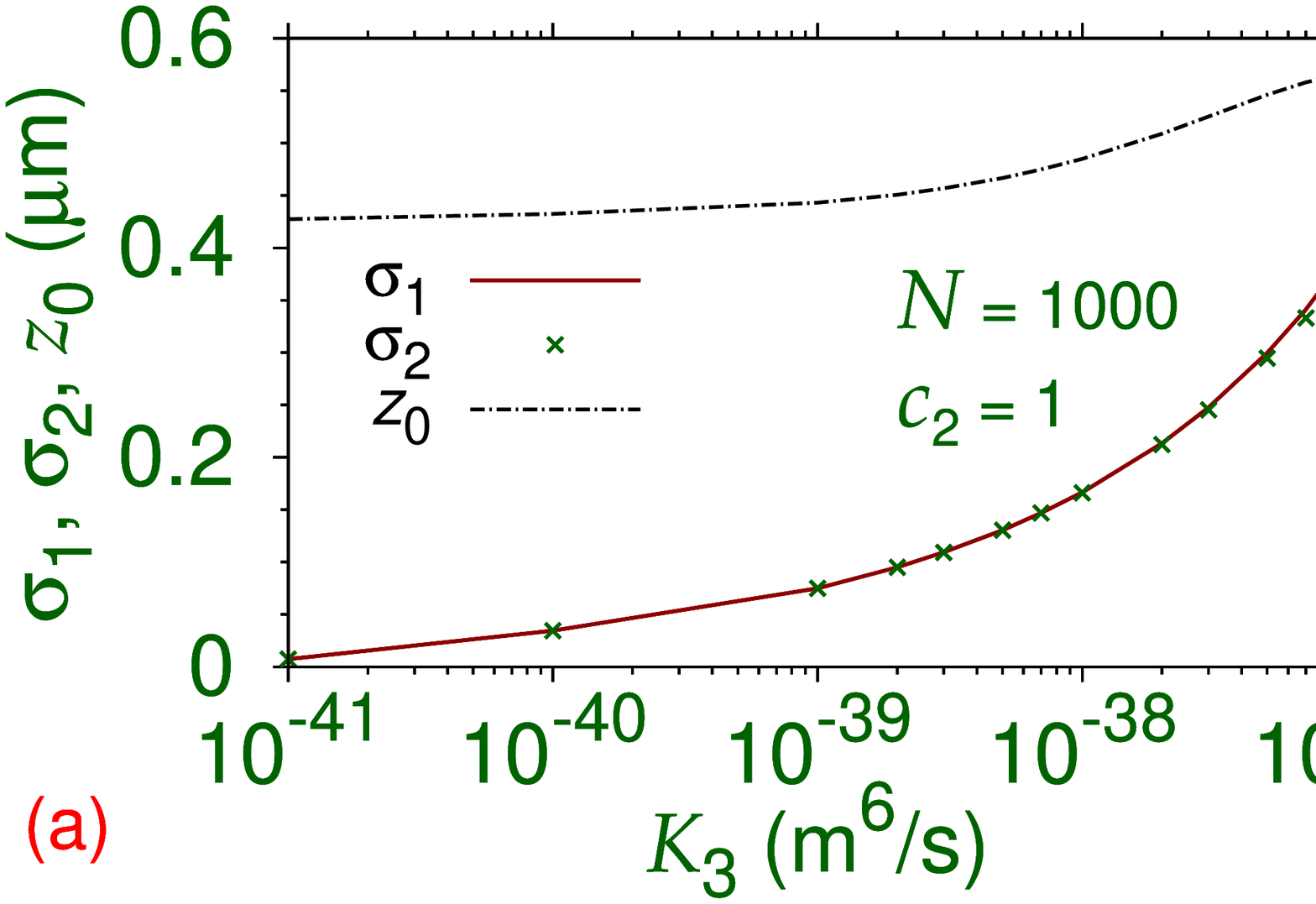}
 \includegraphics[width=.45\linewidth,clip]{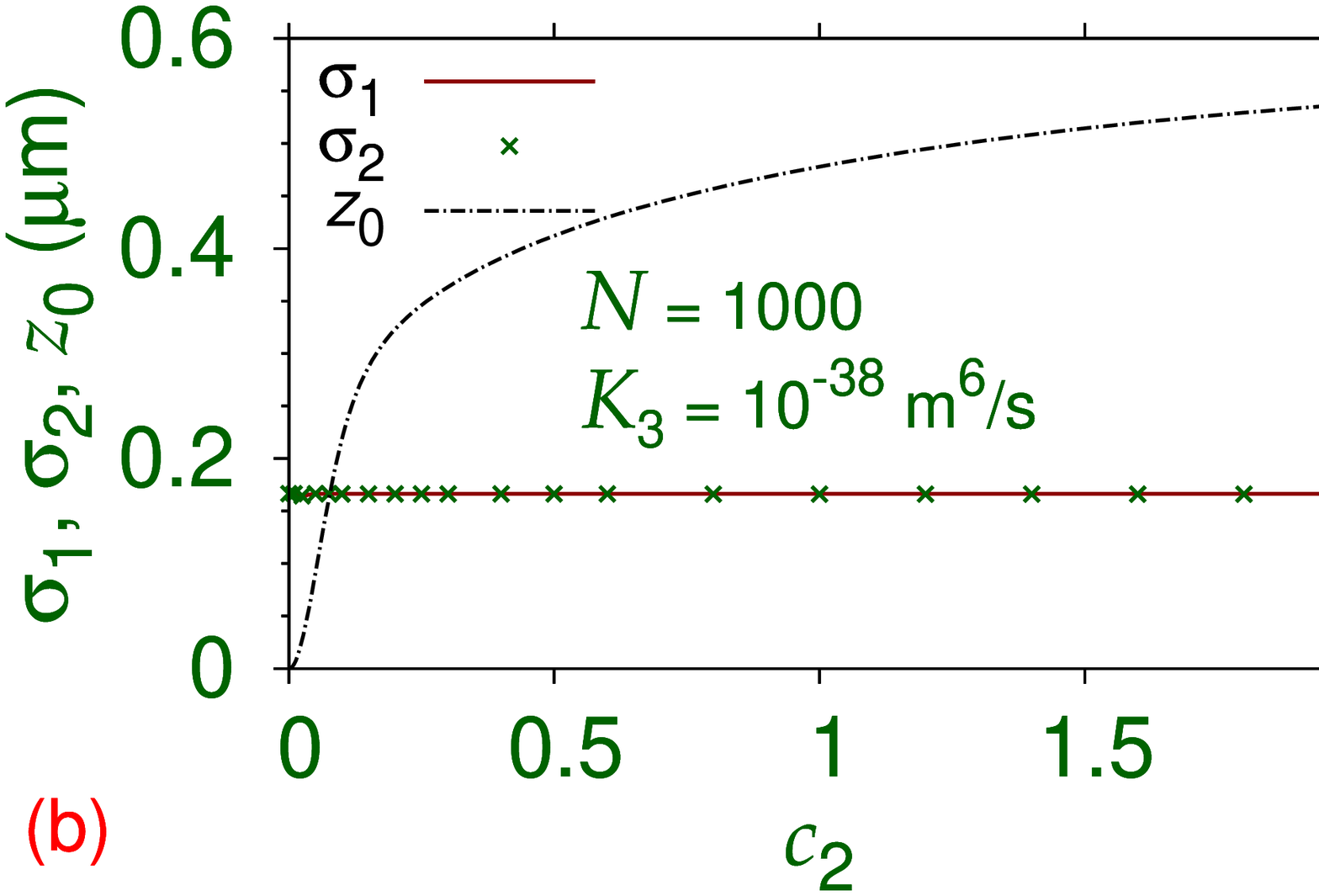} 
 
\caption{   Variational parameters $\sigma_1, \sigma_2, z_0$, from a minimization of the Lagrangian (\ref{eq6}), (a) versus $K_3$ for  $c_2=1$ and (b) versus $c_2$ for   $K_3=10^{-38}$ m$^6$/s. The other parameters are $N =1000, a =-27.4a_0, c_1 =1, \gamma =20.$
}\label{fig1} 

\end{center}

\end{figure}

 We study the SSB states in the 1D double-well potential (\ref{dw})  from a minimization of the variational  Lagrangian (\ref{eq6}). This determines the variational widths $\sigma_1, \sigma_2$ and the parameter $z_0$ which is a measure of symmetry breaking. We consider $N=1000$ $^7$Li atoms.  In Fig. \ref{fig1}(a) we plot the variational parameters $\sigma_1, \sigma_2$ and $z_0$ versus the three-body interaction term $K_3$ for parameters $c_1=c_2=1, \gamma =20$ in 
Eq. (\ref{dw}).  In Fig. \ref{fig1}(b) we show the same parameters versus $c_2$ for three-body interaction $K_3= 10^{-38}$ m$^6$/s and $c_1=1.$   We find that for any non-zero $c_2$, $z_0$ is non zero: a non-zero $c_2$ with $c_1=1$ in Eq. (\ref{dw}) represents a double well and  a non-zero $z_0$ signals SSB. Hence, a symmetry breaking will take place for the weakest possible 1D double-well potential.

 \begin{figure}[!b]
\begin{center}
 \includegraphics[trim=10 0 10 0,clip,width=.45\linewidth]{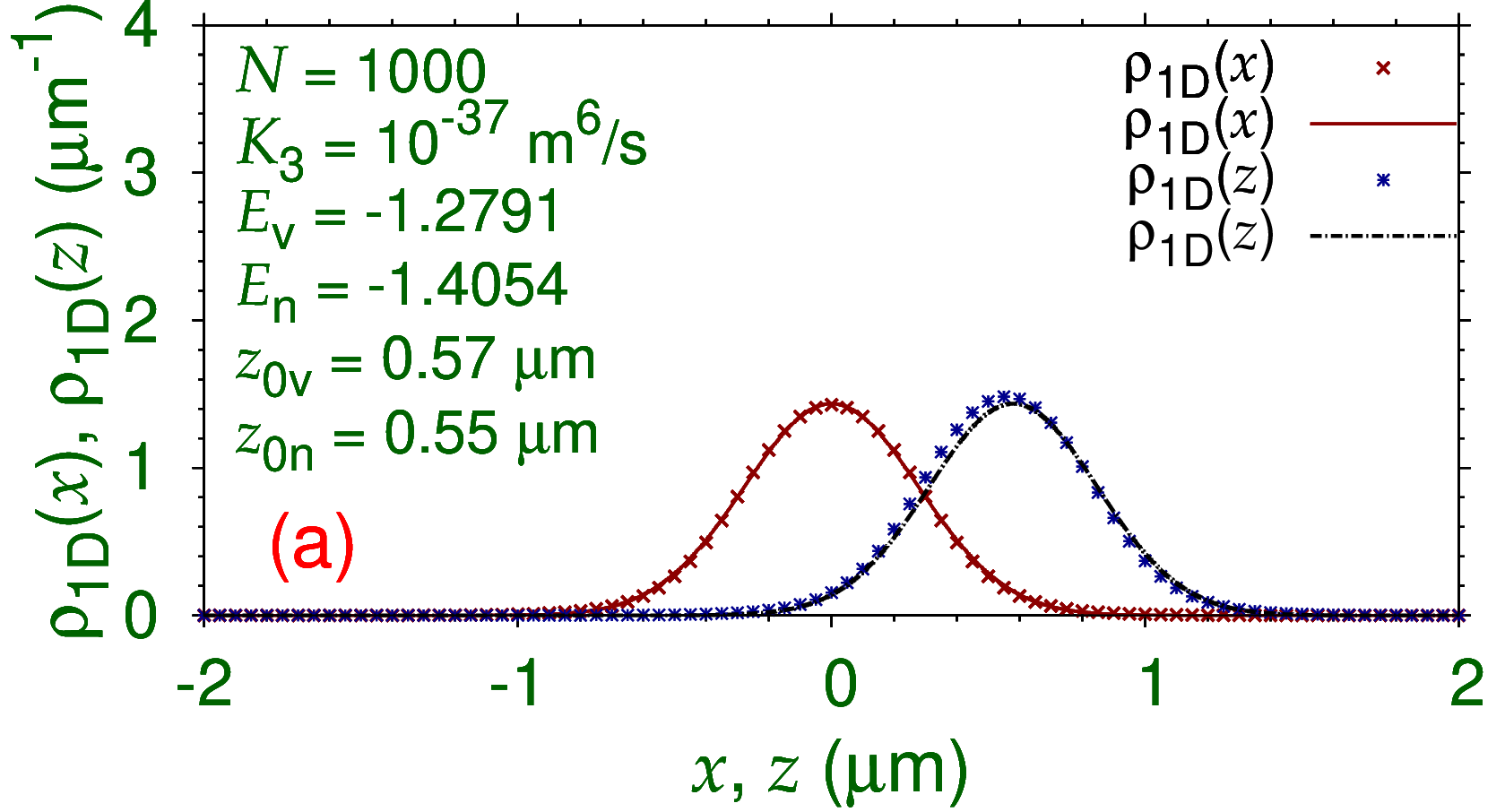}
 \includegraphics[trim=10 0 10 0,clip,width=.45\linewidth]{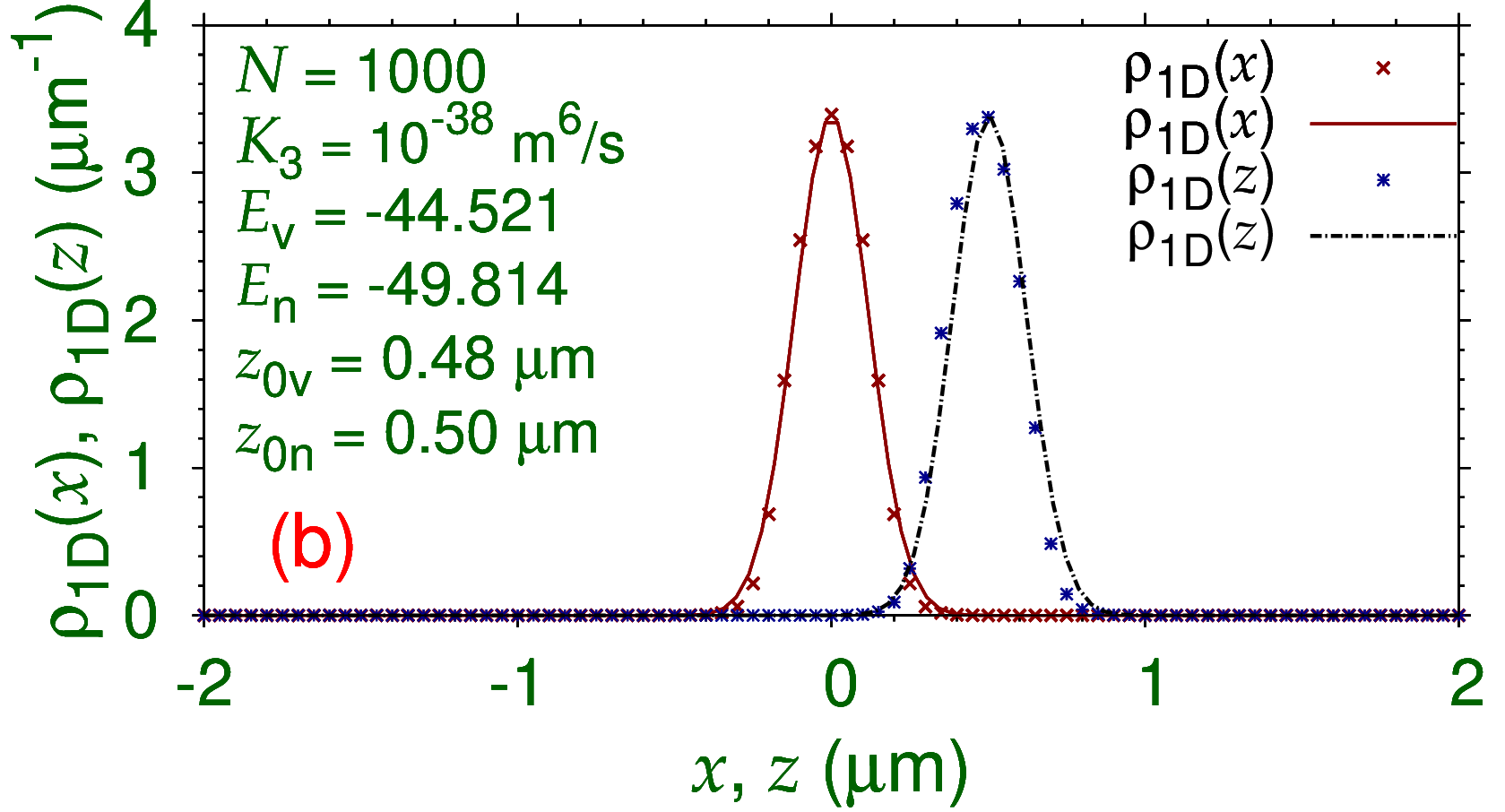}
 \includegraphics[trim=10 0 10 0,clip,width=.45\linewidth]{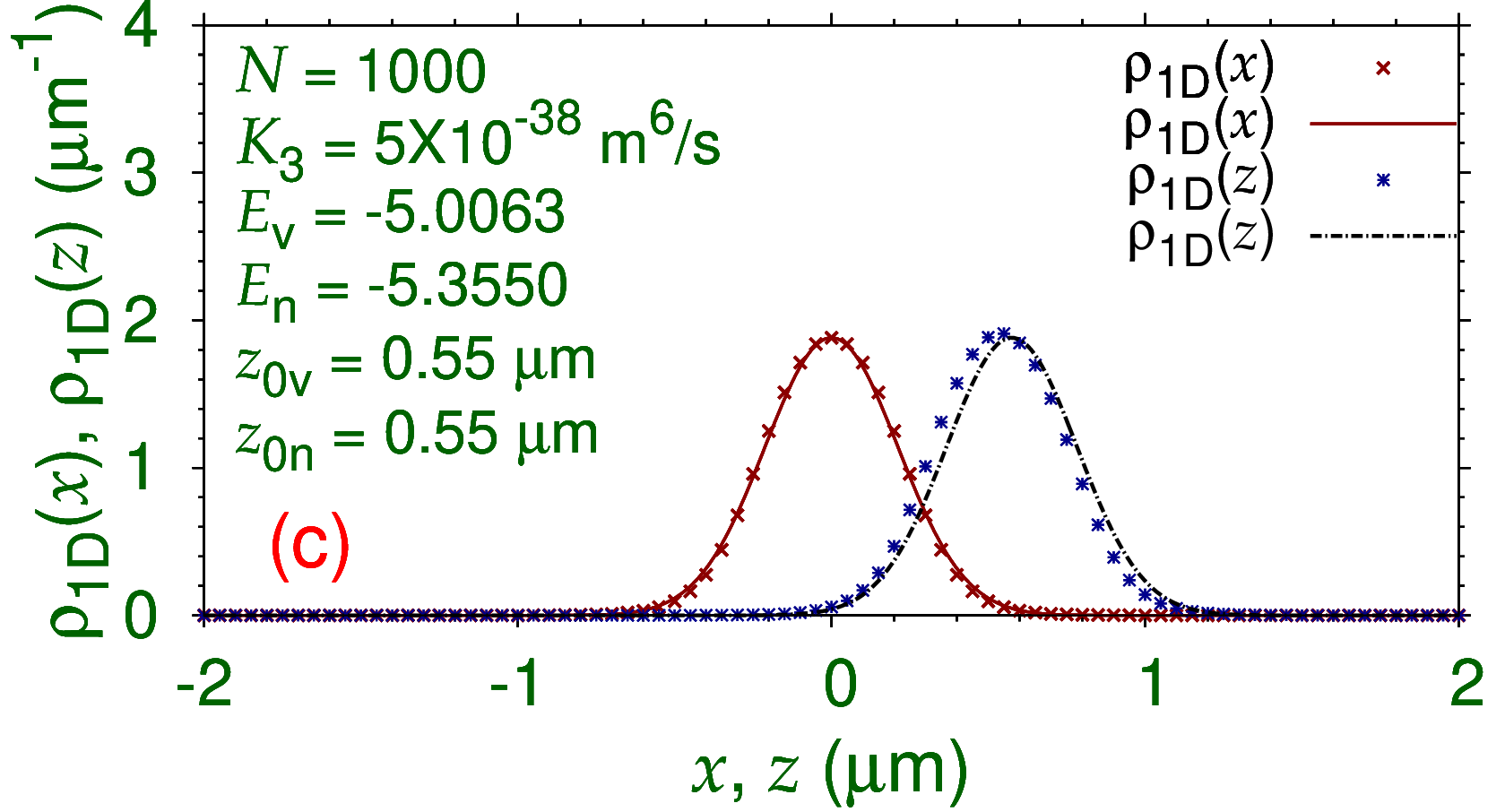}
 \includegraphics[trim=10 0 10 0,clip,width=.45\linewidth]{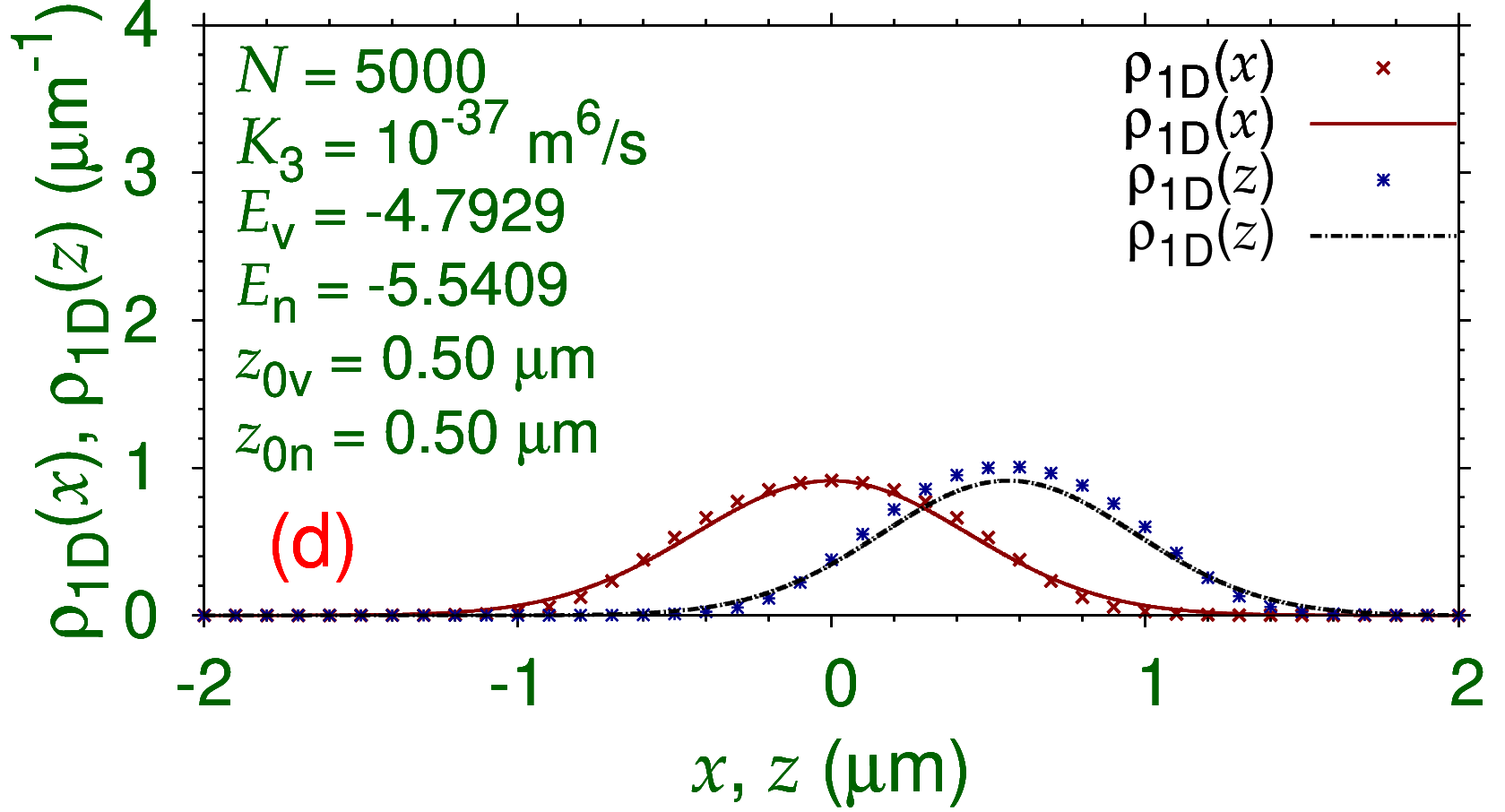}
 
\caption{ Numerical  (points) and  variational (line) 1D reduced  densities $\rho_{1D}(x)$ and $\rho_{1D}(z)$ of SSB quantum balls  of
 $N=1000$ $^7$Li atoms 
and   different three-body term $K_3$ in the double-well trap (\ref{dw}) with $c_1=c_2=1, \gamma=20$. The exhibited numerical (n) and variational (v)  energies per atom given by Eq. (\ref{eq6}) are in units of $9.57\times 10^{-31}$ J. The peak of the density  $\rho_{1D}(z)$ at $z_0$  and its  asymmetric distribution is reasonably well represented by the symmetric Gaussian form of the variational approximation. 
}\label{fig2} 

\end{center}

\end{figure} 

To study the density distribution of the quantum balls,  we define  
reduced 1D  densities  by 
\begin{align}
\rho_{\mathrm{1D}}(x) &= \int dz dy |\phi({\bf r})|^2, \\
\rho_{\mathrm{1D}}(z) &= \int dx dy |\phi({\bf r})|^2.
%\rho_{\mathrm{2D}}(x,y)&= \int dt |\phi({\bf r})|^2.
\end{align} 
The 3D numerical simulation of the GP equation is much more complicated and time consuming compared to the variational approximation considered above. However, to validate the variational findings, for example in Figs. \ref{fig1},  
a comparison to actual numerical results is called for, which we undertake next. The result of this investigation is  illustrated in Figs. \ref{fig2}, where we compare the reduced 1D densities for several SSB bound states along the  $x$ and $z$ axes as obtained  from variational approximation and numerical solution of the GP equation with the double well (\ref{dw})  with parameters $c_1=c_2=1, \gamma=20$. The variational and numerical energies per atom, as given by   Lagrangian (\ref{eq6}),  are also shown in the respective plots. For a fixed number of atoms, $N=1000$, the quantum ball is smaller in size (compact)   for a small three-body term $K_3$.  For a fixed three-body term $K_3$, the quantum ball is more compact for a small number of atoms.  The SSB is also explicit in Figs. \ref{fig2}: the reduced density along $z$ direction is asymmetric and does not have any symmetry around $z=0$ or around $z=z_0$ $-$ the point of density maximum. The SSB quantum balls of Figs. \ref{fig2} are created in one of the  doubly-degenerate states centered at $z=z_0$, the other degenerate state is located at 
$z=-z_0$. The wave functions of these two degenerate states are $z$-parity images of each other: 
$\phi_1(x,y,z)= \phi_2(x,y,-z)$. Considering that the reduced 1D density $\rho_{1D}(z)$ is not symmetric around $z=z_0$ implying a non-Gaussian $z$ profile of the quantum ball, the agreement between the variational and numerical densities is good.

\begin{figure}[!t]

\begin{center}

\includegraphics[width=.245\linewidth,clip]{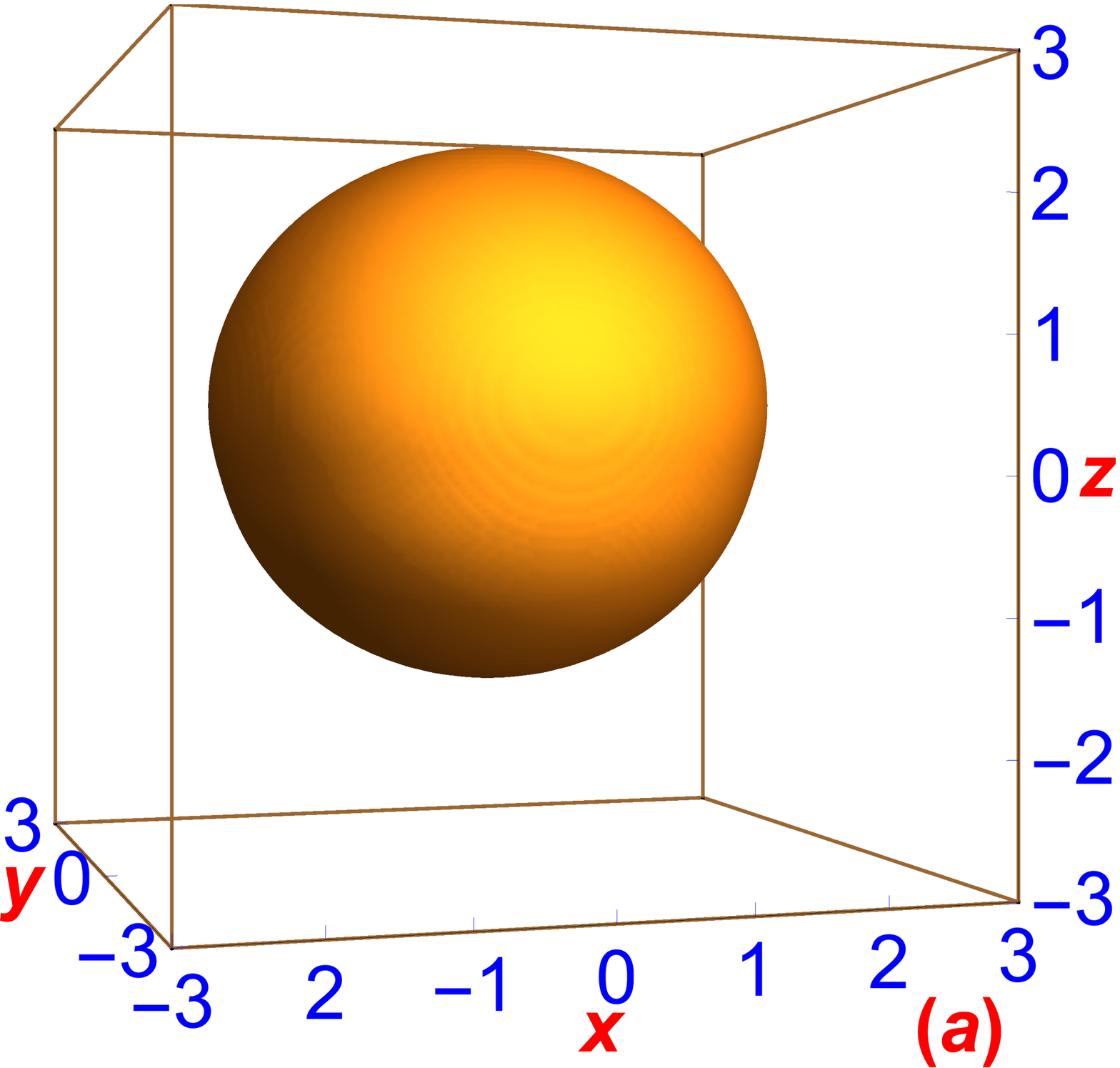}
\includegraphics[width=.245\linewidth,clip]{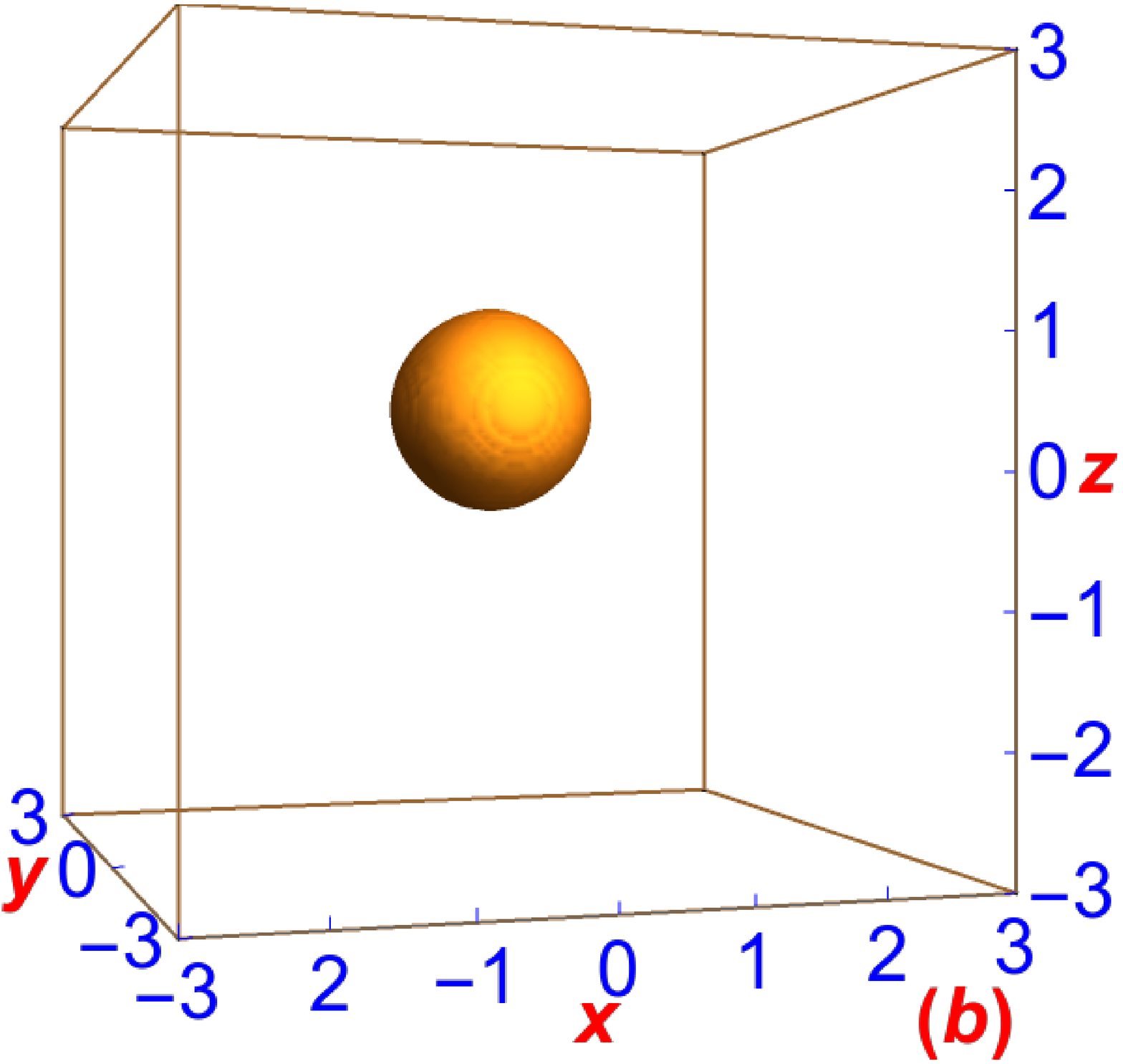}
\includegraphics[width=.245\linewidth,clip]{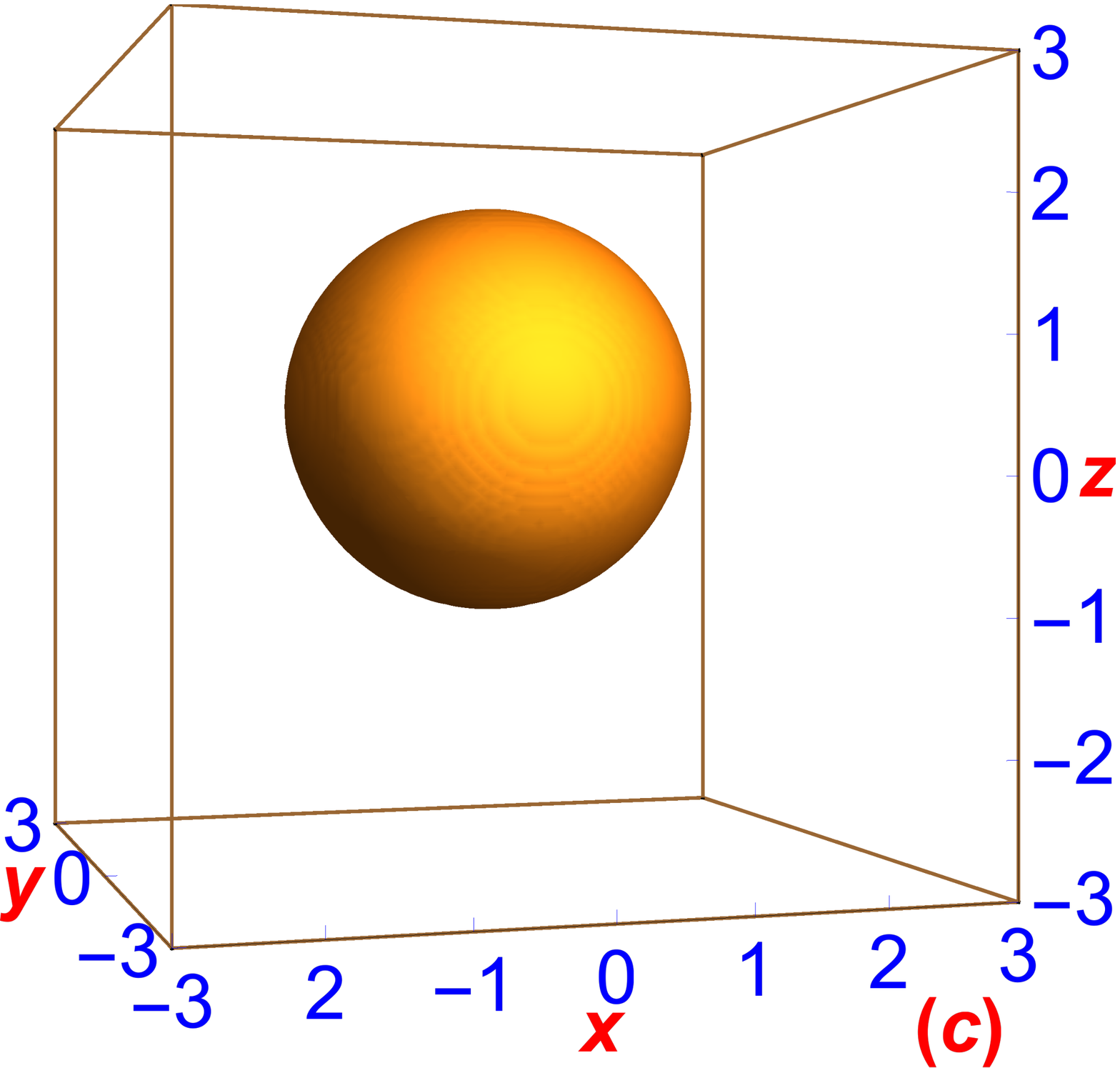}
\includegraphics[width=.245\linewidth,clip]{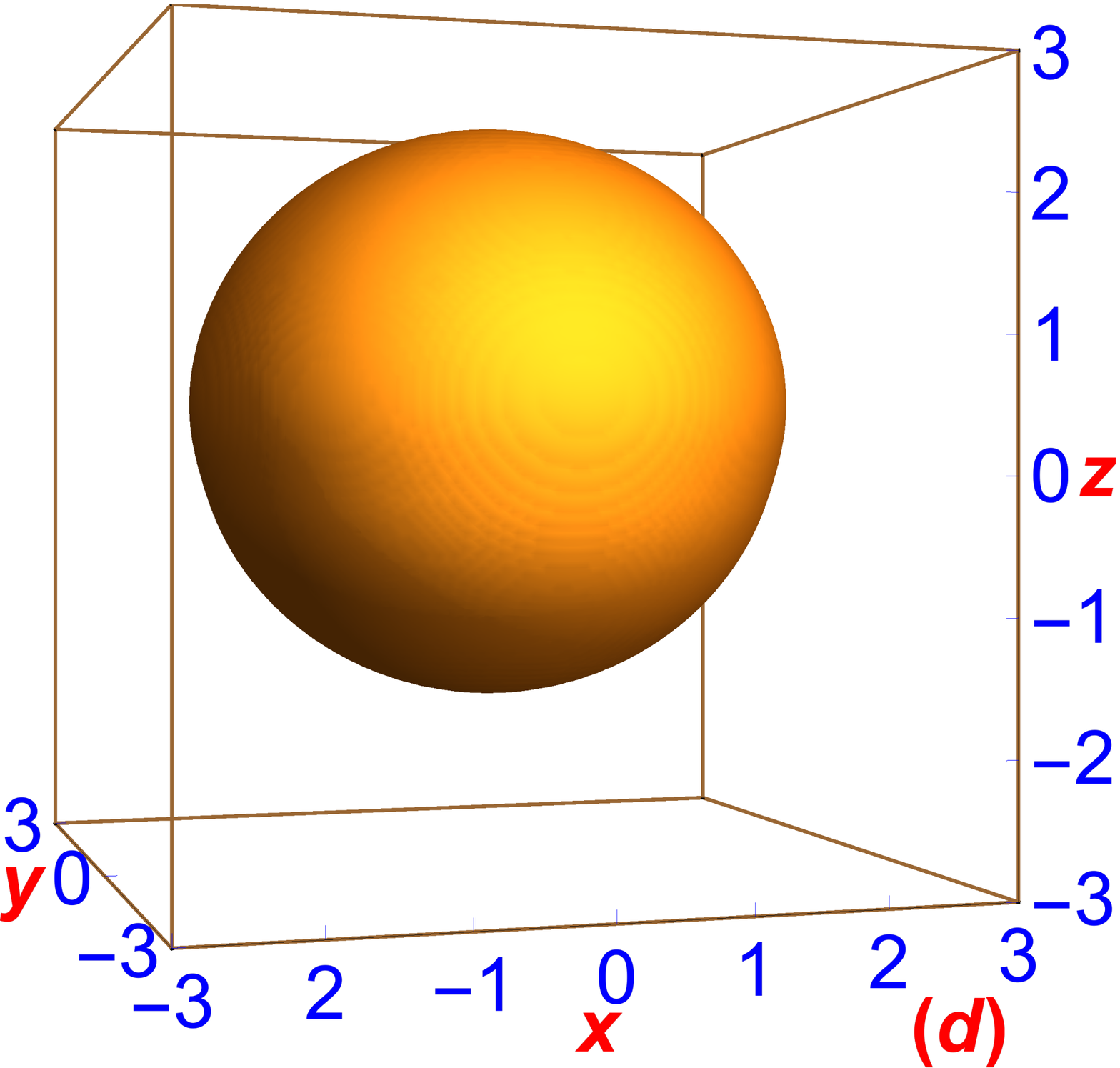}
 
\caption{ 
Three-dimensional  isodensity contour of SSB quantum balls ($N|\phi({\bf r})|^2$) in 1D double-well potential (\ref{dw})
with $c_1=c_2=1, \gamma=20$. The parameters in (a)-(d) are the same as in Figs. \ref{fig2}(a)-(d), respectively:
    (a) $N=1000, K_3= 10^{-37}$ m$^6$/s, (b) $N=1000, K_3= \times 10^{-38}$ m$^6$/s,
(c) $N=1000, K_3= 5\times 10^{-38}$ m$^6$/s, (d) $N=5000, K_3=3\times  10^{-37}$ m$^6$/s. The density of atoms on the contour is $10^9$ atoms/cm$^3$. The units of $x,y,z$ are $\mu$m.
}\label{fig3} \end{center}

\end{figure}

We now consider the  3D profile of the SSB quantum ball in the 1D double-well potential. In Figs. \ref{fig3}(a)-(d), we plot the 3D isodensity contour ($N|\phi({\bf r})|^2$) of the quantum balls exhibited in  Figs. \ref{fig2}(a)-(d), respectively. In Figs. \ref{fig3} we see that as a result of SSB the ball is physically displaced along the $z$ axis slightly  in the positive $z$ direction, which is also implicit in Figs. \ref{fig2}.  The displaced quantum ball is not spherical, but  slightly deformed in the $z$ direction. The free quantum ball in the absence of the double-well potential is spherical.

\begin{figure}[!b]

\begin{center}
 \includegraphics[width=.45\linewidth,clip]{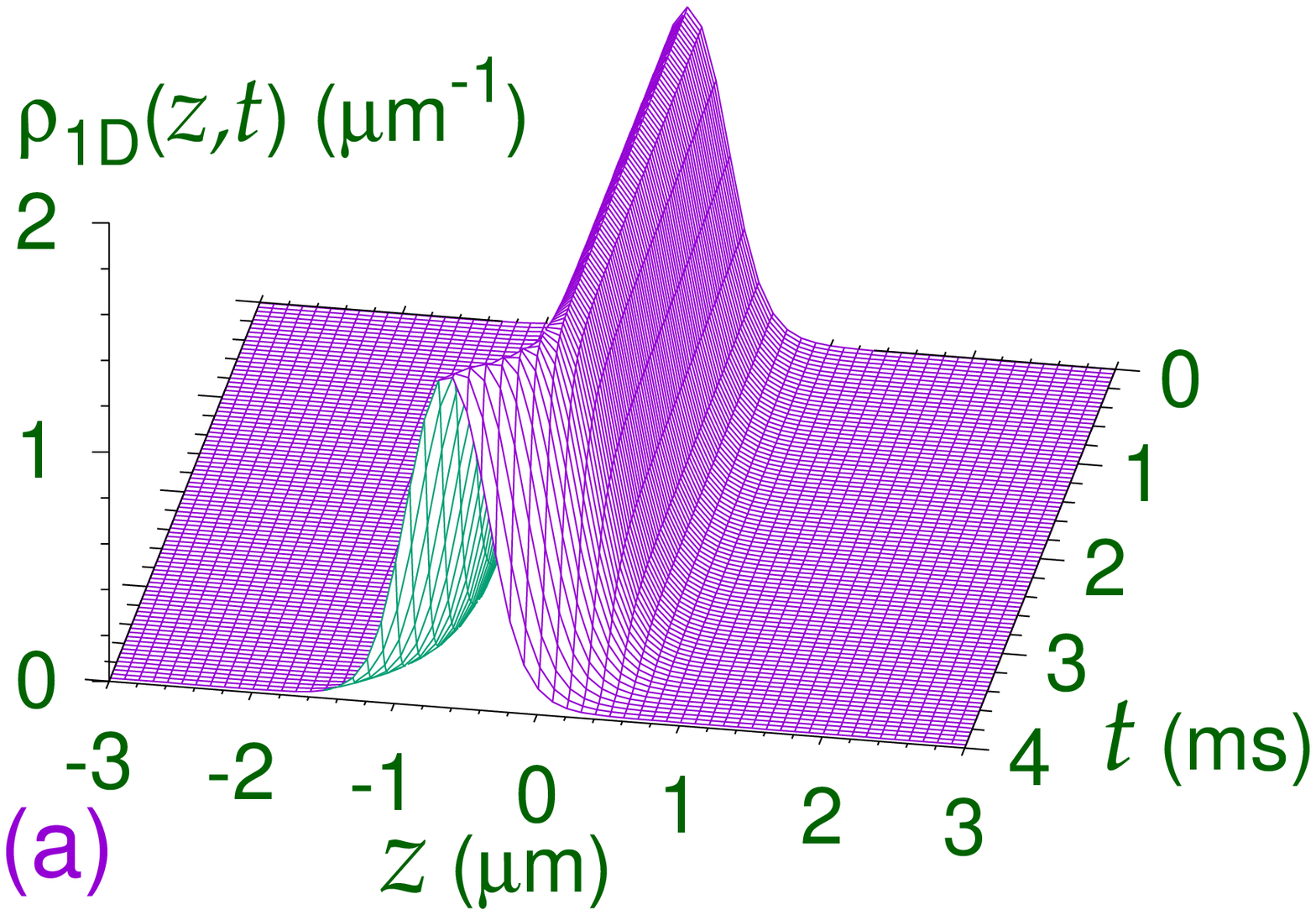}
 \includegraphics[width=.45\linewidth,clip]{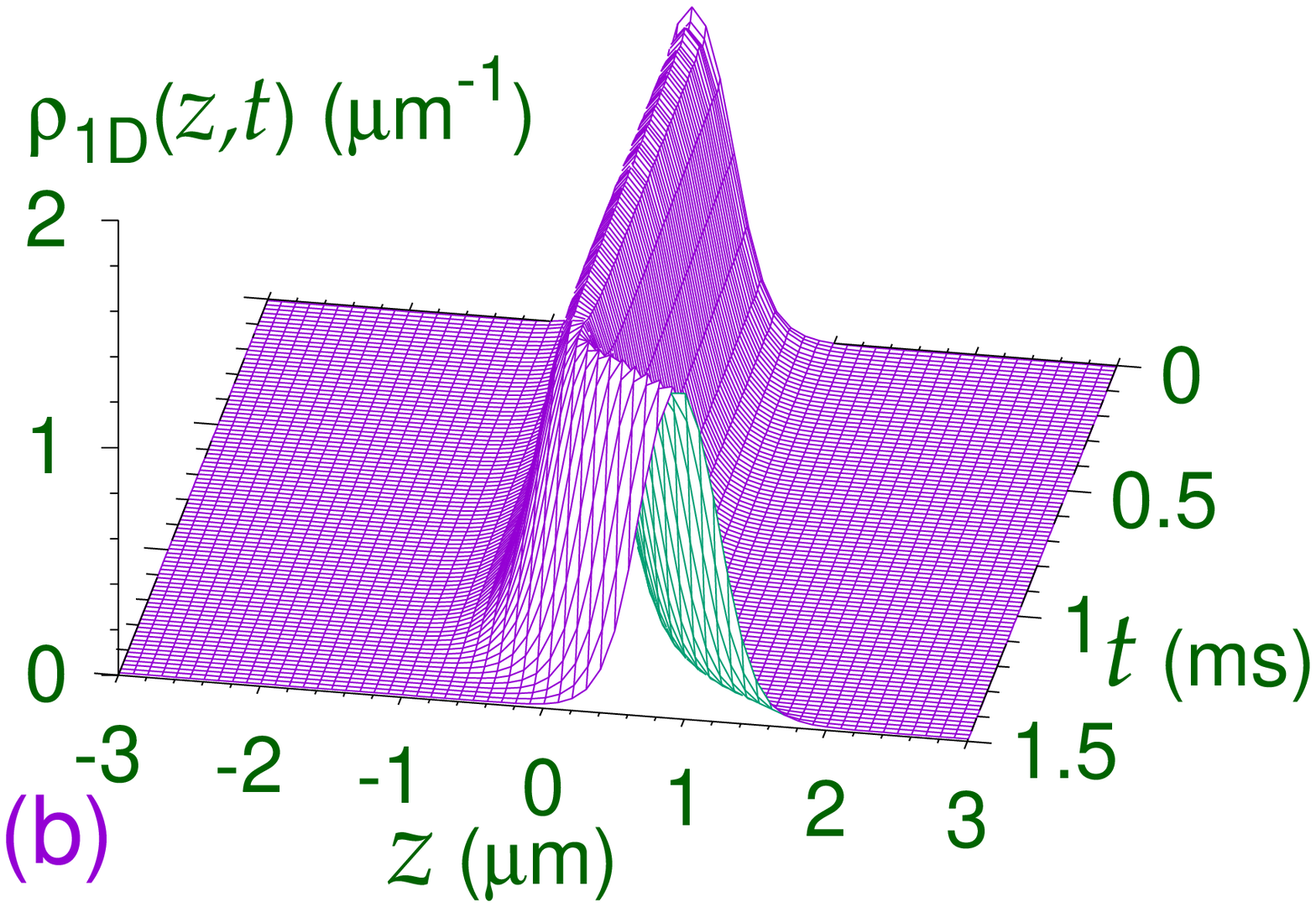} 
 
\caption{  Spontaneous symmetry breaking of a quantum ball made of  $N=1000$ $^7$Li atoms with $a=-27.4a_0$ and $K_3=10^{-37}$ m$^6$/s by real-rime simulation through a plot of 
1D density $\rho_{1D}(z,t)$ versus $z$ and $t$.  (a) A self-bound quantum ball is initially placed at the top of a hill potential $c_2 e^{-\gamma z^2}$ ($c_2=0.05, \gamma=20$)  to study the SSB dynamics. (b) A quantum ball created in the harmonic well $z^2/2$ is placed at the top of the 
hill at ${\bf r}=0$ of the double well $z^2/2 + c_2 e^{-\gamma z^2}$ ($c_2=0.4, \gamma =20$) for the SSB dynamics.
}\label{fig4} 

\end{center}

\end{figure}

To demonstrate the dynamical transition of a quantum ball from a parity-symmetric to a SSB state,  we consider a Gaussian hill potential 
$c_2e^{-\gamma z^2}$  with $c_1=0$ in Eq. (\ref{eq2}). Like a classical ball, the quantum ball placed at the top of a hill  will slide down the hill spontaneously breaking the symmetry. We consider a self-bound quantum ball with $N=1000$ $^7$Li atoms with $K_3=10^{-37}$ m$^6$/s as obtained by solving Eq. (\ref{eq4}) with $c_1=c_2=0$ by imaginary-time propagation.     The quantum ball so obtained is placed at the top of the hill ${\bf r}=0$ and the dynamics studied by solving Eq. (\ref{eq4}) with $c_1=0, c_2=0.05, \gamma=20$.     The quantum ball stays at the top of the hill in unstable equilibrium for some time, but eventually   slides down the hill away from the position of unstable equilibrium spontaneously breaking the symmetry. The interval of time for SSB to start is large when the height of the hill $c_2$ is small  and its width large and vice versa.   This is next demonstrated by real-time simulation for a small $c_2=0.05$. The SSB  is illustrated in Fig \ref{fig4} (a), where we plot the 1D density      
 $
\rho_{\mathrm{1D}}(z,t)$
versus $z$ and $t$ during this dynamics. At about $t\approx 3$ ms, the 1D density of the quantum ball is deviated from the central position indicating the motion of the ball down the hill. 
Once the motion is started the ball will move away from the hill reducing the energy of the system.

Although, a simple hill potential demonstrates SSB, it is not very convenient for studying a 
controlled dynamics in a confined space. To study the dynamics of a SSB state of the quantum ball in a controlled way, we consider next the  double-well potential of Eq. (\ref{eq4}): $c_1 =1, c_2 \ne 0$.  First, by imaginary-time simulation we calculate the ground state of the quantum ball  with 
    $N=1000$ and $K_3=10^{-37}$ m$^6$/s in the single harmonic well:  $c_1 =1, c_2 = 0$. 
   The quantum ball so obtained is placed at the top of the hill ${\bf r}=0$ in the double-well potential and the dynamics studied by solving Eq. (\ref{eq4})   with $c_1=1, c_2=0.4, \gamma=20$.   The SSB dynamics is illustrated in Fig. \ref{fig4} (b), where we again plot the 1D density $\rho_{\mathrm{1D}}(z,t)$ versus $z$ and $t$. At about $t\approx 1$ ms the SSB is manifested and the quantum ball slides down the hill of the double well. However, different from Fig. \ref{fig4}(a), after sliding down the hill the quantum ball now remains confined in space due to the infinite harmonic trap of the double well. In both cases $-$ Figs. \ref{fig4} (a) and (b) 
$-$ there is no preferred direction of sliding from the top of the hill; it is decided by the  
development of numeric in real-time simulation.

Usual Josephson oscillation in cold atoms considers a repulsive trapped BEC through  a narrow barrier. Different from that scenario,    
next we study the Josephson oscillation of the  attractive 3D quantum ball through the narrow barrier  of  the 1D  double-well potential in the $z$ direction with no trap in the perpendicular directions $x$ and $y$.  To  this end, we first solve the GP equation by imaginary-time propagation  to obtain the doubly-degenerate SSB stationary state of the quantum ball in the presence of the double-well potential along the $z$ axis, which we use subsequently in the study of Josephson oscillation. This SSB stationary state is displaced from the center at $z=0$.  The  Josephson oscillation is started by giving the pre-calculated SSB  stationary state in the double-well potential  a displacement $\delta$ along the positive  $z$ axis.  A positive $\delta$ denotes a displacement along the  positive $z$ axis  away from the center of the trap, whereas a negative  $\delta$ denotes a displacement along the negative $z$ axis towards the center of the trap.  
The resultant dynamics is studied by real-time simulation of the GP equation (\ref{eq2}) with the displaced quantum ball in the initial state.   To quantify the   Josephson oscillation we consider the dynamics of population imbalance $S(t)$
\begin{equation}\label{st}
S(t) \equiv \frac{N_1(t)-N_2(t)}{N_1(t)+N_2(t)},
\end{equation}
where $N_1(t)$ and $N_2(t)$ are the number of atoms at time $t$ for $z>0$ and $z<0$, respectively.

For an illustration of Josephson oscillation  and self trapping, we consider the SSB quantum ball shown in Fig. \ref{fig2}(c) with $N=1000, K_3=5\times 10^{-38}$ m$^6$/s. The Josephson oscillation dynamics in this case is illustrated first for positive $\delta$ (initial displacement away from the trap center $z=0$) in Fig. \ref{fig5} through plots of $S(t)$ versus $t$ for  $\delta =+0.6 $ $\mu$m,  +0.55 $\mu$m, +0.2 $\mu$m, +0.05 $\mu$m.
The initial SSB   of Fig. \ref{fig2}(c) with $S(0)= 0.86$ has $93\%$ of atoms in the first well with $z> 0$. For smaller displacements, $\delta =+0.05 $  $\mu$m and  +0.2 $\mu$m, the oscillation of $S(t)$ for solely positive values in Fig. \ref{fig5} indicates  self-trapping of most of the atoms  in the first well: $z>0$.  However, for larger initial displacement, e.g.,   $\delta =+0.55 $  $\mu$m, the oscillation of $S(t)$ covering both positive and negative values indicates Josephson oscillation with most of the atoms oscillating between the two wells. In this case $S(t)$ oscillates between the limiting values $+0.98$ and $-0.3$ indicating that the percentage of atoms in the second well, $z<0$, varies between $1\%$ and $65\%$.  This is a case of asymmetric Josephson oscillation, the asymmetry is introduced due to the SSB of the   initial state located in the first well:  $z>0$.  In the case of Josephson oscillation in a trapped repulsive  BEC,  the initial state is parity symmetric and so is the Josephson oscillation \cite{st1,st1a,st1b,st1c}. 
For a slightly larger initial displacement, e.g.,   $\delta \ge +0.6 $  $\mu$m, $S(t)$ oscillates between the limits $\pm 1$ indicating a  transfer
of all atoms from one well to another and {\it vice versa} during a symmetric Josephson oscillation.

The Josephson oscillation dynamics for negative $\delta $ values (displacement towards the center of the trap at $z=0$) of the SSB quantum ball of Fig. \ref{fig2}(c) is studied next. For small values of initial displacement, $|\delta |= 0.05$ $\mu$m, $0.2$ $\mu$m, the population imbalance $S(t)$ is always positive indicating a self trapping of most atoms in the first well: $z>0$. However, for a medium initial displacement,  $|\delta |= 0.5$ $\mu$m, a significant percentage of atoms could be transferred to the second well during Josephson oscillation dynamics, while $S(t)$ oscillates between positive and negative values $+0.97$ and $-0.3$,  indicating that the percentage 

\begin{twocolumn}

 \begin{figure}[!t]

\begin{center}
\includegraphics[trim = 0mm 0mm 0mm 0mm, clip,width=.85\linewidth]{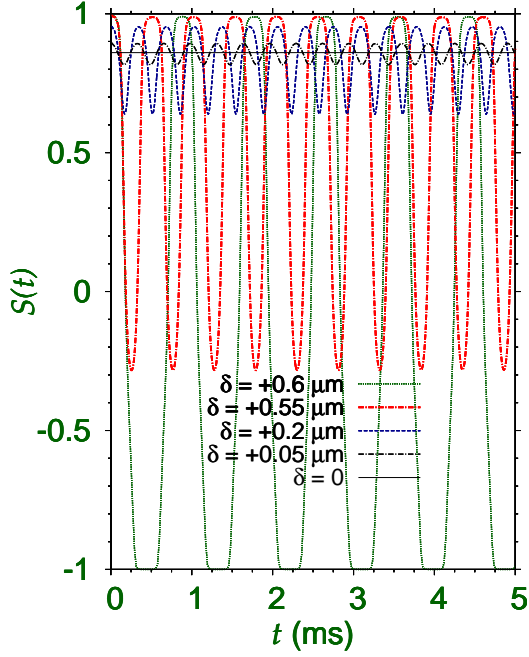}
 
\caption{   Josephson oscillation and self-trapping of the SSB quantum ball of Fig. \ref{fig2}(c) ($N=1000, K_3=5\times  10^{-38}$ m$^6$/s)
 from a dynamics of population imbalance between the two wells. The dynamics is obtained by a real-time evolution of the GP equation (\ref{eq2})  after giving an initial displacement $\delta$ along the positive $z$ axis  to  the SSB stationary state. The positive $\delta $ values indicate a displacement of the initial state away from the trap center at $z=0$.  For smaller initial displacements, ($\delta =+0.05$ $\mu$m, $+0.2$ $\mu$m) the quantum ball is predominantly in the $z>0$ domain denoted by an average  positive $S(t)$ indicating self-trapping in the first well, whereas for larger displacements $(\delta =+0.55$ $\mu$m) a Josephson oscillation is initiated with a tunneling of most atoms between the two wells. For larger displacements ($\delta \ge +0.6$ $\mu$m) a symmetric     Josephson oscillation takes place with a tunneling of all atoms  between the two wells.
}\label{fig5} \end{center}

\end{figure}

 \begin{figure}[!t]

\begin{center}
\includegraphics[trim = 0mm 0mm 0mm 0mm, clip,width=.85\linewidth]{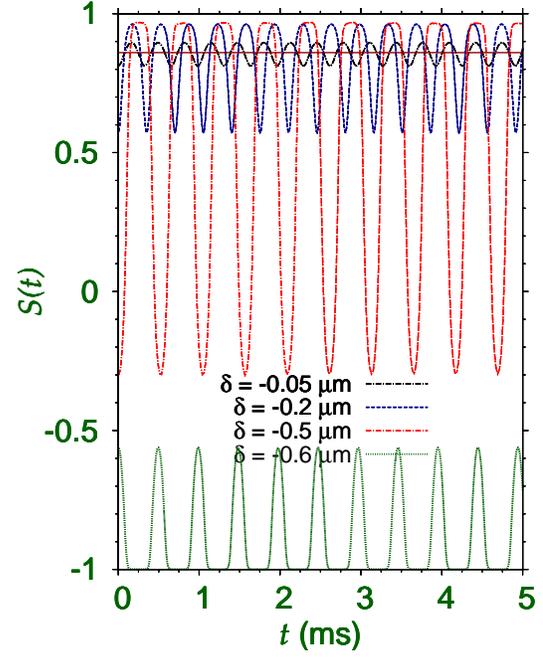}

\caption{  Josephson oscillation and self-trapping of the SSB quantum ball of Fig. \ref{fig2}(c) 
 from a dynamics of population imbalance between the two wells. Refer to Fig. \ref{fig5} and the text for a complete description. The negative $\delta $ values indicate a displacement of the initial state towards  the the trap center at $z=0$.  For smaller initial displacements, 
($|\delta| =+0.05$ $\mu$m, $+0.2$ $\mu$m) the quantum ball is predominantly in the $z>0$ domain denoted by an average  positive $S(t)$ indicating self-trapping in the first well, whereas for larger displacements $(|\delta| =+0.5$ $\mu$m) a Josephson oscillation is initiated with a tunneling of most atoms between the two wells. For larger displacements ($|\delta| = +0.55$ $\mu$m), a permanent self trapping in the second well takes place indicated by negative $S(t)$ vales.  
}\label{fig6} \end{center}

\end{figure}
 
\noindent 
of atoms in the second well, $z<0$, varies between $2\%$ and $65\%$. For a larger initial displacement
$|\delta |= 0.6$ $\mu$m, most atoms of the quantum ball enters the second well ($z<0$) at $t=0$
 with $S(0)\approx -0.55$ and   the quantum ball is accommodated in the 
position of the second  of the doubly-degenerate SSB states in the second well and oscillates around this position leading to negative $S(t)$ values  indicating a permanent self-trapping in the second well.
We find that the dynamics for $\delta =+0.05$ $\mu$m, $+0.2$ $\mu$m and $+0.55$  $\mu$m is surprisingly similar to that for $\delta =-0.05 $ $\mu$m, $-0.2$ $\mu$m and $-0.5$ $\mu$m, respectively.
 For small displacements away from the center, 
viz. Fig. \ref{fig5}, or towards the center, viz. Fig. \ref{fig6}, the quantum ball oscillates around the
position of equilibrium, thus leading to similar dynamics in Figs. \ref{fig5} and \ref{fig6}.
 However, larger displacements towards the center of the trap takes the quantum ball in the initial state close to the position of equilibrium of the second degenerate state at $z<0$ and the quantum ball starts oscillating around this position, viz. $\delta=-0.6$ $\mu$m
in Fig. \ref{fig6}. For a larger initial displacement away from the trap,   viz. $\delta=+0.6$ $\mu$m  in Fig. \ref{fig5}, the quantum ball reaches  the position of equilibrium of the second degenerate state at $z<0$ with a large kinetic energy. Consequently, it cannot be trapped in this position and executes a symmetric Josephson oscillation through the Gaussian barrier at the center of the double well. Even larger  initial displacement towards the center of the trap takes the quantum ball deep into the second well past the position of equilibrium of the second degenerate state at $z<0$
and essentially the same dynamics as illustrated in Figs. \ref{fig5} and \ref{fig6}  emerges, however, with $S(t)$ changed to $-S(t)$ due to the $z$-parity symmetry of the problem.

\end{twocolumn}

\begin{onecolumn}

 The reduced 1D density in the $z$ direction  $\rho_{1D}(z)$ of the SSB quantum-ball  is not symmetric around the $z$ peak at $z=z_0$.  This is already explicit in Fig. \ref{fig6}. The $z$ peak of the SSB quantum ball of Fig. \ref{fig2}(c) is located at $z\equiv z_0 \approx 0.55$. If we give a displacement of this quantum ball  through $\delta =-0.5$
  $\mu$m and $\delta =-0.6$  $\mu$m towards center, respectively, the quantum ball will come to the symmetric positions  $z\approx \pm 0.05$   $\mu$m. As the Hamiltonian is $z$-parity symmetric, these two $z$-symmetric initial configurations, should lead to similar oscillation dynamics (one being parity image of other), if the density   $\rho_{1D}(z)$ were symmetric around the peak at $z=z_0$.  However, the vastly different asymmetric nature of oscillation dynamics for $\delta =-0.5$  $\mu$m and $\delta =-0.6$  $\mu$m in Fig. \ref{fig6} is due to  the  asymmetric  nature of the peak in density  
 $\rho_{1D}(z)$.

Next we illustrate in Fig. \ref{fig7} the appearance of the symmetric (SJO) and asymmetric (AJO) Josephson 
 oscillations, and
self-trapping ST for $z>0$ and $z<0$  of a 3D quantum ball 
in a  double-well potential along the $z$ axis
 in a phase-plot in the 
  (a) $\delta-c_2$ plane and (b) $\delta-N$ plane. The initial SSB quantum ball in the double-well potential is located in $z>0$. These plots are not symmetric around $\delta=0$. For large positive $\delta$ (displacement away from $z=0$)
we have symmetric Josephson oscillation. However, for large negative $\delta$ (displacement towards $z=0$) 
we have self-trapping in the $z<0$ domain. For small $|\delta|$ we have self trapping in the $z>0$ domain.
The transition from one regime to another is sharp, e.g., for a small change in the model parameter 
$\delta$, one can have transition from a symmetric Josephson oscillation to an
 asymmetric Josephson oscillation, then to self-trapping, etc.   
The self-trapping corresponds to a  disbalance of time-averaged atom population in the two wells of the double-well potential or a non-zero time-averaged  $\langle S(t)\rangle$ of Eq. (\ref{st}): $ \langle S(t)\rangle=0$ for the parameter domain of symmetric Josephson oscillation and  $ \langle S(t)\rangle\ne 0$ for all other domains illustrated in Fig. \ref{fig7}, viz. Figs. \ref{fig5} and \ref{fig6}.

 \begin{figure}[!t]

\begin{center}
\includegraphics[trim = 0mm 10mm 0mm 0mm, clip,width=.49\linewidth]{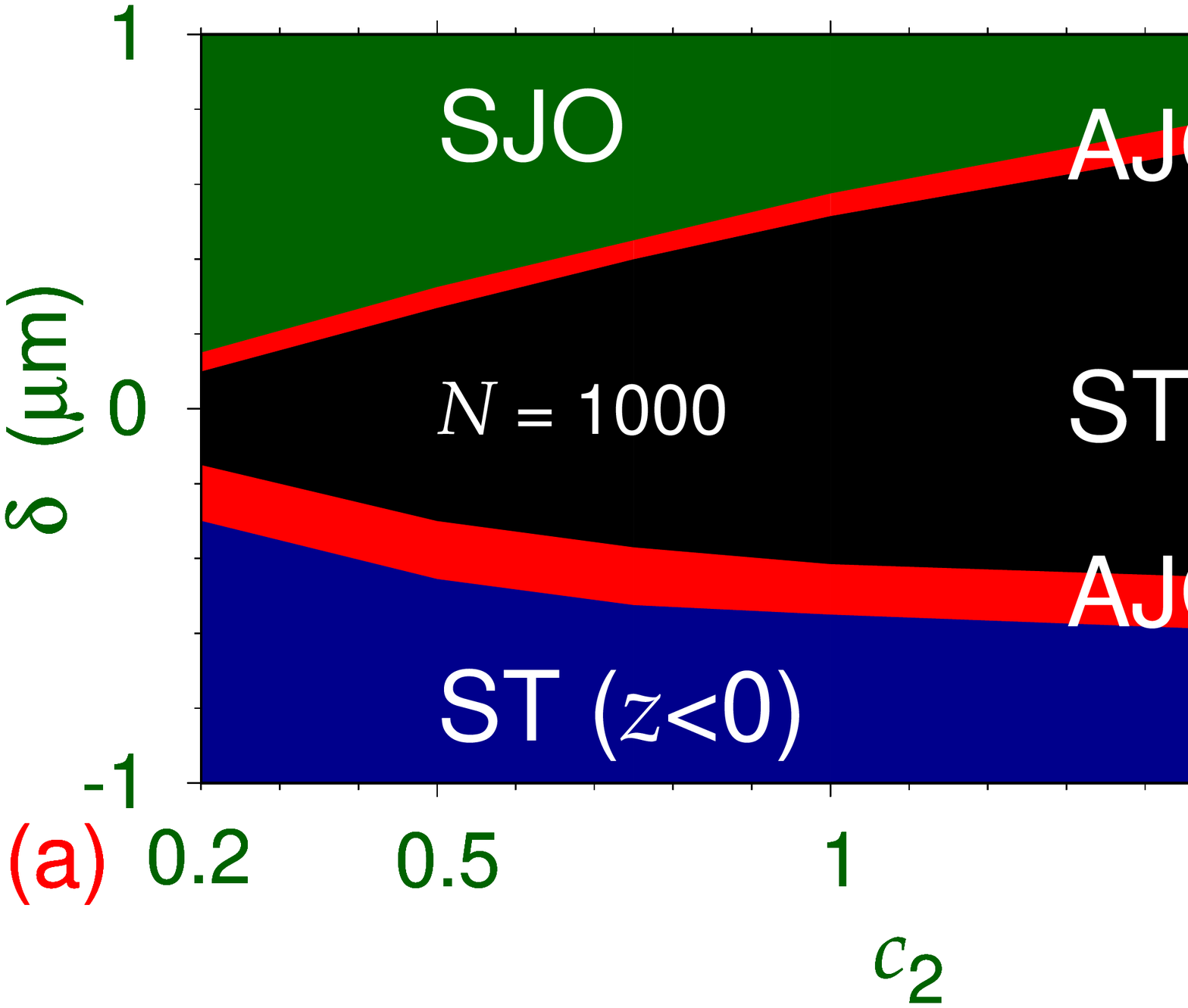}
 \includegraphics[trim = 0mm 0mm 0mm 0mm, clip,width=.49\linewidth]{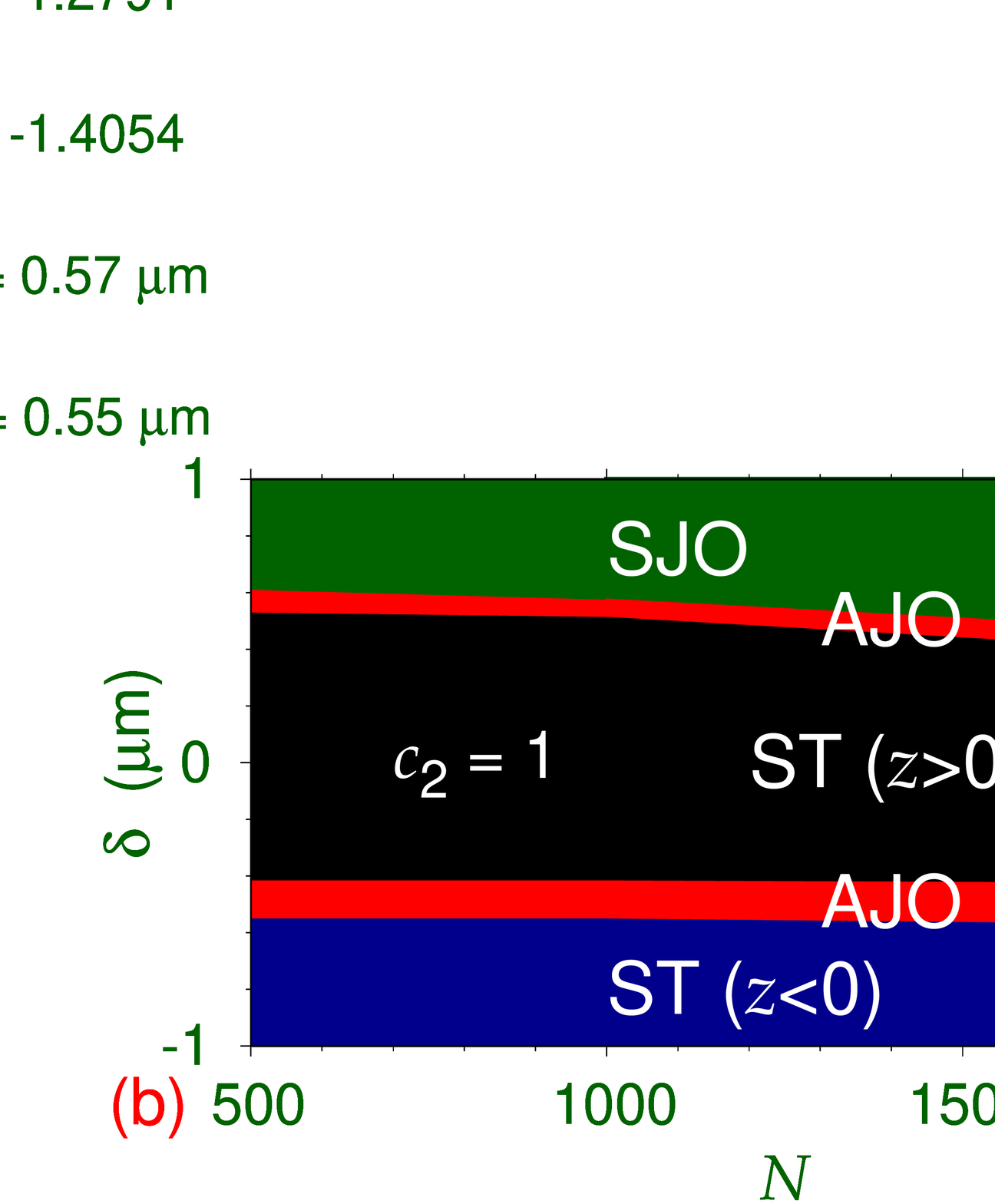}
  
\caption{  Phase plot in the (a) $\delta-c_2$ plane  ($N=1000$) and (b) $\delta-N$  plane  ($c_2=1$) illustrating the symmetric (SJO, Green) and asymmetric (AJO, Red) Josephson oscillation, and self-trapping (ST) for $z>0$ (Black) and $z<0$ (Blue) of a 3D quantum ball in a double-well potential along the $z$ axis.  The fixed parameters of the quantum ball and the double-well potential are  $K_3= 5\times 10^{-38}$ m$^6$/s, $a=-27.4a_0,$ $c_1=1$,  $\gamma=20$.   
}\label{fig7} \end{center}

\end{figure}

\section{Methods}
 
\label{IV}
 
The 3D GP  equation (\ref{eq2})
is generally solved by   the split-step Crank-Nicolson \cite{CPCa,CPCb,CPCc} and Fourier spectral \cite{jpb}
methods.
 The split-step 
Crank-Nicolson method   
  in Cartesian coordinates is employed in the present study.  We use a  space  
  ($\bf r$ $=\{x,y,z\}$)  step of  $0.05$ $\mu$m $ \sim$ $0.025$ $\mu$m,  a  time  step of  $ 0.0005t_0$ ms $  \sim$ $0.00025t_0$  ms \cite{CPCa,CPCb,CPCc} and  the number of  discretization points $192 \sim 256$ in each of $x,y $, and $z$ directions. 
There are different C and FORTRAN programs for solving the GP equations \cite{CPCa,CPCb,CPC1a,CPC1b,CPC1c,CPC1d}
and one should use an appropriate one. 
{We use both imaginary- and real-time propagation \cite{CPCa,CPCb,CPCc} for the numerical solution of the 3D GP equation. The imaginary-time propagation is used   to find the stationary state and the real-time propagation is used in the study of  dynamics employing the initial stationary profile obtained by the imaginary-time propagation.     } In the imaginary-time propagation the initial  state was taken as  in Eq. (\ref{eq3}) with the parameters obtained from the variational solution   (\ref{eq7}).

\section{Discussion}

\label{V}

To summarize,
we demonstrate the formation of SSB (spontaneous symmetry-broken) doubly-degenerate  states of  a 3D quantum ball bound by attractive 
two-body and repulsive three-body interactions in   a  1D double-well  and in a Gaussian potential along the $z$ direction employing a variational approximation  and a numerical solution of the 3D GP equation. The doubly-degenerate states are symmetrically located at $z=\pm z_0$.  
The  double-well  and Gaussian potentials have no effect on the binding of the quantum ball  but are required for 
symmetry breaking.  We also study Josephson oscillation and self-trapping of the 3D quantum ball
in the 1D double-well potential along the $z$ axis. An oscillation of a quantum ball  in the first well ($z>0$) is initiated   
by giving a displacement to it  in the $z$ direction    and the subsequent dynamics is studied to investigate Josephson oscillation and self-trapping. For a small initial displacement towards or away from the center of the trap ($z=0$), one has oscillation of the quantum ball mostly in the first well indicating self-trapping in the first well. For a large initial displacement away 
from the center, one encounters symmetric Josephson oscillation of the quantum ball between the two wells. For a similar displacement towards the center, a self-trapping of the quantum ball in the second well ($z<0$) is realized. For a medium displacement towards the center or away from the center, one has an asymmetric Josephson oscillation of the quantum ball between the two wells.  The relatively large asymmetry of the  Josephson oscillation is due to asymmetric profile of the quantum ball on two sides of the density peak in the $z$ direction at  $z= z_0$ and is an earmark of spontaneous symmetry breaking. 

In this paper we considered a self-bound quantum ball under the action of   attractive two-body and repulsive three-body interactions. There are other suggestions to make a self-bound quantum ball \cite{other,me1}. The principal conclusions of the present study should also be applicable to 
quantum balls prepared in a different fashion. 
Apart from being of theoretical interest in SSB in nonlinear dynamics, the present study is also of phenomenological interest.  
 To  observe the asymmetric Josephson oscillation reported in this paper, one should  consider a very weak trap in the $x$ and $y$ directions so as to localize the quantum ball in an experiment and consider a double-well potential in the $z$ direction. Such an experiment seems possible in the future.

\section*{Acknowledgements}
We thank the Funda\c c\~ao de Amparo 
\`a
Pesquisa do Estado de S\~ao Paulo (Brazil)
(Project:  2012/00451-0)
  and  the
Conselho Nacional de Desenvolvimento   Cient\'ifico e Tecnol\'ogico (Brazil) (Project: 303280/2014-0) for 
support.

\section*{Additional information}

%To include, in this order: \textbf{Accession codes} (where applicable); \textbf{Competing financial interests} (mandatory statement). 

The author declares no competing financial interests.

\end{onecolumn}

\begin{thebibliography}{99}

\bibitem{symb}% A. Trenkwalder,	G. Spagnolli,	G. Semeghini,	S. Coop,	M. Landini,	P. Castilho,	L. Pezzè,	G. Modugno,	M. Inguscio,	A. Smerzi,	
%and M. Fattori,     
 Trenkwalder, A. {\it et al.}, Quantum phase transitions with parity-symmetry breaking and hysteresis,
{\it Nature Phys.}
{\bf    12},
    826–829
    (2016).
 
 \bibitem{m1}Zegadlo, K. B.,   Dror,  N.,Trippenbach,  M.,   Karpierz, M. A. \&
 Malomed,  B. A., 
Spontaneous symmetry breaking of self-trapped and leaky modes in quasi-double-well potentials,
{\it Phys. Rev. A} {\bf 93}, 023644 (2016).
\bibitem{m2} 
 Shamriz, E.,  Dror, N. \&  Malomed, B. A., 
Spontaneous symmetry breaking in a split potential box,
{\it Phys. Rev. E} {\bf 94}, 022211 (2016);
\bibitem{m2a}
 Mayteevarunyoo, T., Malomed, B. A. \& Dong, G., 
Spontaneous symmetry breaking in a nonlinear double-well structure,
{\it
Phys. Rev. A} {\bf 78}, 053601 (2008).

\bibitem{s1a}  Gautam, S. \& Adhikari, S. K., Spontaneous symmetry breaking in a spin-orbit-coupled f=2 spinor condensate,
{\it Phys. Rev. A} {\bf 91}, 013624 (2015).
\bibitem{s1b}
Adhikari, S. K.,
Demixing and symmetry breaking in binary dipolar Bose-Einstein-condensate solitons,
{\it Phys. Rev. A} {\bf 89},  013630 (2014).
\bibitem{s1c}
 Cheng,   Y.  \&  Adhikari,  S. K., Symmetry breaking in a localized interacting binary Bose-Einstein condensate in a bichromatic optical lattice,
{\it Phys. Rev. A} {\bf 81}, 023620 (2010).
%\bibitem{s1d}
%S. K. Adhikari, Laser Phys. Lett. {\bf 7}, 824 (2010).

\bibitem{fermion} Adhikari, S. K.,  Malomed, B. A., Salasnich,  L. \&  Toigo, F., 
Spontaneous symmetry breaking of Bose-Fermi mixtures in double-well potentials,
{\it Phys. Rev. A} {\bf 81}, 053630 (2010).
 
 
 
\bibitem{m3a}Heil, T.,  Fischer, I., Els\"asser,  W., Mulet, J. \& Mirasso,  C. R.,
Chaos synchronization and spontaneous symmetry-breaking in symmetrically delay-coupled semiconductor lasers,
{\it Phys. Rev. Lett.} {\bf 86}, 795-798 (2001).
\bibitem{m3b}
%P. Hamel, S. Haddadi, F. Raineri, P. Monnier, G. Beaudoin, I. Sagnes, A. Levenson, and A. M. Yacomotti,  
 Hamel P. {\it et al.},
Spontaneous mirror-symmetry breaking in coupled photonic-crystal nanolasers,
{\it Nature Phot.} {\bf 9}, 311-315 (2015).
%\bibitem{m3c} E. B. Davies,  Commun. Math. Phys. {\bf 64}, 191  (1979).
%\bibitem{m3d} J. C. Eilbeck, P. S. Lomdahl, and A. C. Scott,  Physica D {\bf 16}, 318 (1985).
\bibitem{m3e} Malomed,  B. A., 
 Symmetry breaking in laser cavities,
{\it Nature Phot.} {\bf 9}, 287-289 (2015).

 
\bibitem{n1a}Sadler, L. E.,  Higbie,  J. M., Leslie,  S. R.,  Vengalattore, M. \&   Stamper-Kurn, D. M.,
Spontaneous symmetry breaking in a quenched ferromagnetic spinor Bose-Einstein condensate,
{\it Nature }  {\bf 443}, 312-315 (2006),
\bibitem{n1b}
Yukalov,  V. I., 
Bose-Einstein condensation and gauge symmetry breaking,
{\it	Laser Phys. Lett.} {\bf 4}, 632-647 (2007).
 \bibitem{n1c} 
%M. Scherer, B.Lucke, J. Peise, O. Topic, G. Gebreyesus, F. Deuretzbacher, W. Ertmer, L. Santos, C. Klempt, and J. J. Arlt,  
 Scherer M.  {\it et al.}, 
Spontaneous symmetry breaking in spinor Bose-Einstein condensates,
{\it Phys. Rev. A}  {\bf 88}, 053624 (2013).
% E. H. Lieb, R. Seiringer, J. Yngvason,  	Rep. Math. Phys. {\bf 59}, 389 (2007); J. I. Kapusta, Phys. Rev. D {\bf 24}, 426 (1981).
 
 
 \bibitem{sol}
 Kivshar, Y. S. \& Malomed, B. A., Dynamics of solitons in nearly integrable systems,
{\it Rev. Mod. Phys.} {\bf 61}, 763-915 (1989).

%F. K. Abdullaev, A. Gammal, A. M. Kamchatnov, and L. Tomio,
%Int. J. Mod. Phys. B {\bf 19}, 3415 (2005).



\bibitem{n2} Salasnich, L.  \& Malomed,   B. A., 
Spontaneous symmetry breaking in linearly coupled disk-shaped Bose-Einstein condensates,
{\it	Molecular Phys.} {\bf 109},  2737-2745 (2011). 



\bibitem{josep1}%  F.S.  Cataliotti,  S.  Burger,  C.  Fort,  P.  Maddaloni,  F.
%Minardi, A. Trombettoni, A. Smerzi, M. Inguscio, 
 Cataliotti,  F. S.  {\it et al.},
Josephson junction arrays with Bose-Einstein condensates,
{\it Science}
{\bf 293}, 843-846 (2001).
 \bibitem{josep1a}
Levy, S.,   Lahoud,  E.,Shomroni,  I. \&   Steinhauer, J.,
The a.c. and d.c. Josephson effects in a Bose-Einstein condensate,
{\it Nature}  {\bf 449}, 579-U8 (2007).
 \bibitem{josep1b}
 %L. J. LeBlanc, A. B. Bardon, J. McKeever, M. H. T. Extavour, D. Jervis, J. H. Thywissen, F. Piazza, and A. Smerzi,
LeBlanc,  L. J. {\it et al.}, Dynamics of a Tunable Superfluid Junction,
{\it Phys. Rev. Lett.} {\bf 106}, 025302 (2011).
 %\bibitem{josep1c}
%M. Albiez, R. Gati, J. Folling, S. Hunsmann, M. Cristiani, and M. K. Oberthaler,
%Phys. Rev. Lett. {\bf 95}, 010402 (2005).
 \bibitem{josep1d}
Adhikari, S. K., 
Mean-field model for Josephson oscillation in a Bose-Einstein condensate on an one-dimensional optical trap,
{\it Eur. Phys. J. D} {\bf 25}, 161-166 (2003).


\bibitem{st1}Raghavan, S.,  Smerzi, A.,  Fantoni,  S., \& Shenoy,  S. R., 
Coherent oscillations between two weakly coupled Bose-Einstein condensates: Josephson effects, pi oscillations, and macroscopic quantum self-trapping,
{\it Phys. Rev. A} {\bf 59}, 620-633 (1999).
% A. Smerzi and S. Raghavan, {\it ibid.} {\bf 61}, 063601 (2000);
\bibitem{st1a}
Xiong, B.,  Gong,  J.-B.,  Pu, H.,  Bao,  W.-Z., \&  Li,   B.-W., 
Symmetry breaking and self-trapping of a dipolar Bose-Einstein condensate in a double-well potential,
{\it  Phys. Rev. A}  {\bf 79}, 013626 (2009).
\bibitem{st1b}
Adhikari, S. K.,
Self-trapping of a dipolar Bose-Einstein condensate in a double well,
{\it Phys. Rev. A} {\bf 89},  043609 (2014).
\bibitem{st1c}
 Ananikian,   D.  \&  Bergeman, T., 
Gross-Pitaevskii equation for Bose particles in a double-well potential: Two-mode models and beyond,
{\it Phys. Rev. A} {\bf 73}, 013604  (2006).
% I. I. Satija, R. Balakrishnan, P. Naudus, J. Heward, M. Edwards, and C. W. Clark,
%Phys. Rev. A {\bf 79}, 033616 (2009); H. M. Cataldo and D. M. Jezek,
%Phys. Rev. A {\bf 90}, 043610 (2014).



 \bibitem{stf}
Adhikari, S. K.,  Lu, Hong, \&   Pu,  Han,
Self-trapping of a Fermi superfluid in a double-well potential in the Bose-Einstein-condensate-unitarity crossove,
{\it Phys. Rev. A} {\bf 80},  063607 (2009). 
 




 
\bibitem{qb1}Adhikari, S. K., 
Statics and dynamics of a self-bound matter-wave quantum ball,
{\it Phys. Rev. A} {\bf 95}, 023606 (2017). 

\bibitem{qb2}Adhikari, S. K.,
Statics and dynamics of a self-bound dipolar matter-wave droplet,
{\it Laser Phys. Lett}. {\bf  14}, 025501 (2017). 

\bibitem{qb3}Adhikari, S. K., 
Elastic collision and molecule formation of spatiotemporal light bullets in a cubic-quintic nonlinear medium,
{\it  Phys. Rev. E} {\bf 94}, 032217 (2016).


 \bibitem{qb4}Adhikari, S. K., Elastic collision and breather formation of spatiotemporal vortex light bullets in a cubic-quintic nonlinear medium,  {\it Laser Phys. Lett.} {\bf  14}, 065402  (2017).
 
 
\bibitem{cq}  Berezhiani, V. I., Skarka,  V.  \& Aleksi\'c,   N. B., 
Dynamics of localized and nonlocalized optical vortex solitons in cubic-quintic nonlinear media,
{\it Phys. Rev. E}
{\bf 64}, 057601 (2001).

%\bibitem{cqa} 
%Liu   B.    \&  He,  X.-D., 
%High pressure effect on the ultrafast energy relaxation rate of LDS698 (C19H23N2O4Cl) in a solution,
%{\it Opt. Express}
%{\bf 19}, 20009 (2010).
%\bibitem{cqb} 
%P. Grelu, J. Soto-Crespo, and N. Akhmediev, Opt. Express,
%{\bf 13}, 9352 (2005).
\bibitem{cqb}
 Aleksi\'c,  N. B.,Skarka,  V.,  Timotijevic, D. V. \&    Gauthier,  D., 
Self-stabilized spatiotemporal dynamics of dissipative light bullets generated from inputs without spherical symmetry in three-dimensional Ginzburg-Landau systems,
{\it Phys.  Rev.  A}
{\bf 75},  061802(R) (2007).
%A.  Kamagate,  Ph.  Grelu,  P.  Tchofo-Dinda,  J.
%M.  Soto-Crespo,  and  N.  Akhmediev,  Phys.  Rev.  E
%{\bf 79}, 026609 (2009); S. Chen, Phys. Rev. A
%{\bf 86}, 033829 (2012);
\bibitem{cqc} Mihalache,  D.{\it et al.}, 
Stable vortex tori in the three-dimensional cubic-quintic Ginzburg-Landau equation,
%D. Mihalache, D. Mazilu, F. Lederer, Y. V. Kartashov, L.-C. Crasovan, L. Torner, and B. A. Malomed, 
{\it Phys. Rev.
Lett.}
{\bf 97},  073904  (2006).
% \bibitem{cqd} J.  M.  Soto-Crespo,  Ph.  Grelu,
%and N. Akhmediev, Opt. Express
%{\bf 14}, 4013 (2006).
% N. N.
%Rozanov, J. Opt. Technol.
%{\bf 76}, 187 (2009); 
 %\bibitem{cqe}
%D. Mihalache,
%D. Mazilu, F. Lederer, H. Leblond, and B. A. Malomed,
%Eur. Phys. J. Special Topics
%{\bf 173}, 245 (2009).



 
  
\bibitem{var}Perez-Garcia, V. M.,   Michinel,  H., Cirac,  J. I., Lewenstein,   M. \& P. Zoller,
%Phys. Rev. Lett.  {\bf 77},   5320 (1996);
Dynamics of Bose-Einstein condensates: Variational solutions of the Gross-Pitaevskii equations,
{\it Phys. Rev. A} {\bf 56}, 1424-1432 (1997).

 


\bibitem{abr}  Abraham, E. R. I., McAlexander,  W. I., Sackett,  C. A. \&  Hulet, R. G., 
Spectroscopic determination of the S-wave scattering length of lithium, 
{\it Phys. Rev. Lett.}  {\bf 74}, 1315-1318 (1995). 

\bibitem{loss} Shotan,    Z.,     Machtey,  O.,       Kokkelmans,     S.,  \&  
Khaykovich,  L.,
Three-Body Recombination at Vanishing Scattering Lengths in an Ultracold Bose Gas,
{\it Phys. Rev. Lett.}
{\bf 113},
053202 (2014).



\bibitem{CPCa} Muruganandam  P. \& Adhikari, S. K., Fortran programs for the time-dependent Gross-Pitaevskii equation in a fully anisotropic trap, 
{\it Comput. Phys.
Commun.} {\bf 180}, 1888-1912 (2009).
  \bibitem{CPCb}
 Vudragovi\'c, D., Vidanovi\'c, I., 
 Bala\v z, A., Muruganandam  P. \& Adhikari, S. K., 
C programs for solving the time-dependent Gross-Pitaevskii equation in a fully anisotropic trap,
{\it Comput.
Phys. Commun.} {\bf 183}, 2021-2025 (2012).
  \bibitem{CPCc} Young-S.,  L. E., Vudragovi\'c, D.,  Muruganandam  P.,  Adhikari, S. K.  \& Bala\v z, A., OpenMP Fortran and C programs for solving the time-dependent Gross-Pitaevskii equation in an anisotropic trap,
{\it Comput. Phys. Commun.} {\bf 204},  209-213 (2016).  

\bibitem{jpb} Muruganandam  P. \& Adhikari, S. K., Bose-Einstein condensation dynamics in three dimensions by the pseudospectral and finite-difference methods, {\it J. Phys. B }
{\bf 36}, 2501-2513     (2003).
 
\bibitem{CPC1a}Satari\'c,  B.  {\it et al.}, Hybrid OpenMP/MPI programs for solving the time-dependent Gross-Pitaevskii equation in a fully anisotropic trap,
%B. Satari\'c, V. Slavni\'c, A. Beli\'c, Antun Bala\v z, P. Muruganandam, and S. K. Adhikari, 
{\it Comput.
Phys. Commun.} {\bf 200}, 411-417 (2016).
 \bibitem{CPC1b}  Loncar, V.     {\it et al.}, CUDA programs for solving the time-dependent dipolar Gross-Pitaevskii equation in an anisotropic trap, 
%V. Loncar, A. Bala\v z, A. Bogojevi\'c, S. Skrbi\'c, P. Muruganandam, and S. K. Adhikari,
{\it Comput.
Phys. Commun.} {\bf 200}, 406-410 (2016).
 \bibitem{CPC1c}
 Loncar, V.     {\it et al.},
OpenMP, OpenMP/MPI, and CUDA/MPI C programs for solving the time-dependent dipolar Gross-Pitaevskii equation,
%V. Loncar, L. E. Young-S., S. Skrbic, P. Muruganandam, S. K. Adhikari, A. Bala\v z, 
{\it  Comput. Phys. Commun.} {\bf 209}, 190-196 (2016).
 \bibitem{CPC1d}Kishor Kumar, R.  {\it et al.},
Fortran and C programs for the time-dependent dipolar Gross-Pitaevskii equation in an anisotropic trap,
%R. Kishor Kumar, L. E. Young-S., D. Vudragovi\'c, Antun Bala\v z, P. Muruganandam, and S. K. Adhikari,  
{\it Comput.
Phys. Commun.} {\bf 195}, 117-128 (2015).


\bibitem{other}
 Maucher, F. {\it et al.}, 
Rydberg-Induced Solitons: Three-Dimensional Self-Trapping of Matter Waves,
{\it Phys. Rev. Lett.} {\bf 106,}  170401 (2011).

\bibitem{me1}  Petrov, D. S.,  
Quantum mechanical stabilization of a collapsing Bose-Bose mixture.
{\it Phys. Rev. Lett.}
{\bf 115}, 155302
(2015).

 



   
\end{thebibliography}
\end{document}